\documentclass[prc,aps,floats,floatfix,twocolumn,showpacs,nofootinbib,%
superscriptaddress]{revtex4}

\usepackage{amsmath}
\usepackage{amssymb}
\usepackage{amsfonts}
\usepackage{graphicx}
\usepackage{tabularx}
\usepackage{multirow}
\usepackage{lscape}
\usepackage{rotating}

\usepackage{pstricks}

\renewcommand{\vec}[1]{{\mathbf #1}}
\newcommand{\etal}{\emph{et~al.}}
\newcommand{\nn}{\nonumber}
\newcommand{\noi}{\noindent}

\newcommand{\vsigma}{\boldsymbol{\sigma}}

\newcommand{\wt}{\widetilde}

\newcommand{\alp}{\alpha}

\newcommand{\wuzI}{W_{1}^{{\rm (0,I)}}}
\newcommand{\wdzI}{W_{2}^{{\rm (0,I)}}}

\newcommand{\wuuI}{W_{1}^{{\rm (1,I)}}}
\newcommand{\wduI}{W_{2}^{{\rm (1,I)}}}

\newcommand{\wwwuzI}{{\wt{\wt{W}}_{1}^{{\rm (0,I)}}}}
\newcommand{\wwwuuzI}{{\wt{\wt{W}}_{1}^{{\rm (1,0,I)}}}}
\newcommand{\wwwuuuI}{{\wt{\wt{W}}_{1}^{{\rm (1,\pm 1,I)}}}}

\newcommand{\bdbt}{\beta_2-\beta_3}
\newcommand{\pbdbt}{\left( \beta_2-\beta_3 \right)}

\newcommand{\chihf}{\chi_{HF}}
\newcommand{\chiz}{\chi_0}
\newcommand{\chid}{\chi_2}
\newcommand{\chiq}{\chi_4}

\newcommand{\chizd}{\chiz^2}
\newcommand{\chidd}{\chid^2}
\newcommand{\cchi}{\chiz \chiq}

\newcommand{\bbK}{\boldsymbol{\mathbb K}}

\newcommand{\bbKzz}{\bbK_{\rm  0,0}}
\newcommand{\bbKzu}{\bbK_{\rm  0,1}}
\newcommand{\bbKuz}{\bbK_{\rm  1,0}}
\newcommand{\bbKuu}{\bbK_{\rm  1,1}}
\newcommand{\bbKmz}{\bbK_{\rm -1,0}}
\newcommand{\bbKum}{\bbK_{\rm  1,-1}}
\newcommand{\bbKmm}{\bbK_{\rm -1,-1}}

\newcommand{\demi}{\frac{1}{2}}
\newcommand{\pdemi}{\tfrac{1}{2}}

\newcommand{\moq}{\frac{m^*\omega}{q}}

\newcommand{\moqd}{\left( \moq \right)^2}

\newcommand{\mkdpi}{\frac{m^* k_F}{2\pi^2}}
\newcommand{\mktpi}{\left( \frac{m^* k_F}{3\pi^2} \right)}

\newcommand{\coka}{\left[ 1 + 3(1-k^2)f(k) \right]}
\newcommand{\mrho}{m^{*} \rho}

\newcommand{\br}{\vec{r}}

\newcommand{\bk}{{\bf k}}

\newcommand{\bq}{{\bf q}}

\newcommand{\la}{\langle \,}
\newcommand{\ra}{\, \rangle}
\newcommand{\q}{\quad}

\newcommand{\be}{\begin{equation}}
\newcommand{\ee}{\end{equation}}
\newcommand{\bqr}{\begin{eqnarray}}
\newcommand{\eqr}{\end{eqnarray}}
\newcommand{\bi}{\begin{itemize}}
\newcommand{\ei}{\end{itemize}}
\newcommand{\bc}{\begin{center}}
\newcommand{\ec}{\end{center}}
\newcommand{\bwt}{\begin{widetext}}
\newcommand{\ewt}{\end{widetext}}

\newcommand{\citeqdot}[1]{Eq.~(\ref{#1})}

\newcommand{\citeqssdot}[3]{Eqs.~(\ref{#1}), (\ref{#2}) and~(\ref{#3})}
\newcommand{\citerefdot}[1]{ref.\cite{#1}}

\newcommand{\citeFigure}[1]{Figure~\ref{#1}}

\newcommand{\citeAppendix}[1]{Appendix~\ref{#1}}

\begin{document}

\title{Nuclear response for the Skyrme effective interaction with zero-range 
       tensor terms. II. Sum rules and instabilities.}


\author{A. Pastore}
\email{pastore@ipnl.in2p3.fr}
\affiliation{Universit\'e de Lyon, F-69003 Lyon, France; Universit\'e Lyon 1,
             43 Bd. du 11 Novembre 1918, F-69622 Villeurbanne cedex, France\\
             CNRS-IN2P3, UMR 5822, Institut de Physique Nucl{\'e}aire de Lyon}

\author{D. Davesne}
\email{davesne@ipnl.in2p3.fr}
\affiliation{Universit\'e de Lyon, F-69003 Lyon, France; Universit\'e Lyon 1,
             43 Bd. du 11 Novembre 1918, F-69622 Villeurbanne cedex, France\\
             CNRS-IN2P3, UMR 5822, Institut de Physique Nucl{\'e}aire de Lyon}

\author{Y. Lallouet}
\affiliation{Universit\'e de Lyon, F-69003 Lyon, France; Universit\'e Lyon 1,
             43 Bd. du 11 Novembre 1918, F-69622 Villeurbanne cedex, France\\
             CNRS-IN2P3, UMR 5822, Institut de Physique Nucl{\'e}aire de Lyon}

\author{M. Martini}
\email{martini@ulb.ac.be}
\affiliation{Institut d'Astronomie et d'Astrophysique, CP-226\\
             Universit{\'e} Libre de Bruxelles,
             B-1050 Bruxelles,
             Belgium}

\author{K. Bennaceur}
\email{bennaceur@ipnl.in2p3.fr}
\affiliation{Universit\'e de Lyon, F-69003 Lyon, France; Universit\'e Lyon 1,
             43 Bd. du 11 Novembre 1918, F-69622 Villeurbanne cedex, France\\
             CNRS-IN2P3, UMR 5822, Institut de Physique Nucl{\'e}aire de Lyon}

\author{J. Meyer}
\email{jmeyer@ipnl.in2p3.fr}
\affiliation{Universit\'e de Lyon, F-69003 Lyon, France; Universit\'e Lyon 1,
             43 Bd. du 11 Novembre 1918, F-69622 Villeurbanne cedex, France\\
             CNRS-IN2P3, UMR 5822, Institut de Physique Nucl{\'e}aire de Lyon}


\begin{abstract}
The formalism of linear response theory for Skyrme forces including tensor terms presented in article~\cite{Davesne09} is generalized for the case of a Skyrme energy density functional in infinite matter.
We  also present analytical results for the odd-power sum rules, with particular attention to the inverse energy weighted sum rule, $M_{-1}$,  as a tool to detect instabilities in Skyrme functionals.

\end{abstract}


 
\date{\today}


\maketitle

%
\section{Introduction}
\label{sect:intro}
%

The Energy Density Functional (EDF) method is a tool of choice to perform systematic calculations
of binding energies and one-body observables in the region of the nuclear chart that ranges from medium
to heavy mass atomic nuclei from drip line to drip line~\cite{RMP}. This effective approach relies on
a limited number of universal parameters, usually fitted on experimental
data (observables)~\cite{Kortelainen10,Kortelainen11} along with properties of infinite nuclear matter (pseudo-observables) extracted from experimental results or derived
from realistic models ~\cite{meyer03}.

In its general formulation, the EDF is the sum of different terms that depend only on products on one-body densities
weighted by coupling constants, which in general can also depend on the local densities themselves.
Although several functionals exist on the market~\cite{Robledo10,Hupin11,Carlsson08}, the most
often used is the one derived from the effective Skyrme interaction~\cite{Per04a}.
Building all possible combinations up to quadratic terms in densities together with the conservation of
some general symmetries (see ref. \cite{Per04a} for a detailed discussion), one obtains 28 free coupling
constants~\cite{Raimondi11}, that can be reduced to 14 imposing that the functional is derived from an
effective force. This  requirement is not only adopted to simplify the optimization procedure used
to determine the values of the coupling constants, but it is also mandatory in order to use methods which
go beyond the mean-field to avoid self-interactions and self-pairing \cite{Lacroix09,Lacroix09b,Lacroix09c}. Due to this additional requirement, it turns out that the
standard Skyrme EDF usually adopted in mean-field calculation is not flexible enough to improve
the level of accuracy in describing  set of available experimental data. For this reason, other terms are
now considered as 3-body \cite{Jeremy11} and tensor forces, for example.

The determination of accurate values for the coupling constants of the Skyrme functional, even in its simplest form,
it is quite an elaborated process since a good predictive power is only possible from a pertinent and well-chosen
set of observables or pseudo-observables. Although this is still possible for the time-even part of the
functional~\cite{Kortelainen11}, there is not yet a consensus on how to proceed in order to constrain the time-odd terms.
Actually, it is not clear at all which observables could be used for this purpose, and the corresponding parameters
are not explicitly constrained, but merely indirectly determined by the time-even part through simple mathematical relations.
For this reason, a vast area can be explored in the parameter space and possibly ends up in some  region of instabilities as discussed further in this article.

One of the first method used to fix some of these terms can be found in the work of Van Giai and Sagawa \cite{Giai81},
where they adjust the Landau parameters on values obtained from realistic forces. Furthermore, from the theory of Landau-Migdal for quasi-particles,
one can derive set of sum-rules for Landau parameters ~\cite{Colo10} that should be fulfilled, otherwise the system could
pass through different phase transitions according to the different spin/isospin channels (for instance ferromagnetic instabilities
in spin channels). The Landau-Migdal approach is valid for quasi-particles interacting near the Fermi surface with transfer momentum that
goes to zero - a situation that corresponds to the so-called \emph{long wavelength} limit. Thus it is not able to predict possible instabilities that occur
at non-zero transferred momentum $\mathbf{q}$, with the appearance of domains with typical size  $\lambda\sim2\pi/q$. The first example of such kind of
instability was encountered and examined in details in an article devoted to the study of effective mass splitting by Lesinski
\emph{et al.}~\cite{Les06a}, in  the scalar-isoscalar channel of the SkP functional.
It has been shown that when performing high-accuracy Hartree-Fock calculations (HF) of doubly-magic nuclei, the system converges 
towards an unphysical configuration where protons are separated from neutrons. This observation has also  been confirmed by RPA
calculations in finite nuclei~\cite{Terasaki06}. Another recent example of instability was found by Hellemans \emph{et al.}~\cite{veerle11}
in the vector channel of several Skyrme functionals. They have performed cranked-HFB calculations in $^{194}$Hg and shown
that for particular values of the time odd coupling constants the system can spontaneously polarize.

To improve the existing functionals, it is therefore mandatory to find a tool which is able to detect these instabilities
in all scalar (vector)/isoscalar (isovector) channels. It has already been demonstrated by Lesinski \emph{et al.}~\cite{Les06a} that
the linear response formalism (LR) applied to the Skyrme energy functional could be used to predict the appearance
of some finite-size instabilities in nuclei. However, only the central part of the Skyrme interaction was taken into account for the building
of the linear response. The same LR formalism for a case of a Skyrme interaction including tensor and spin-orbit term was 
studied by  Davesne \emph{et al.}~\cite{Davesne09}, hereafter denoted as article I. In the present article, we extend the formalism
of article I by expressing our results in terms of coupling constant of a general Skyrme functional. The main goal is actually
to investigate the role of odd-power sum rules~\cite{Lipparini} and show that they can be used as a simple and very powerful tool to
detect instabilities in infinite symmetric matter. In particular we give the explicit expression of the inverse sum rule $M_{-1}$ and
demonstrate that a pole in the response function corresponds to a zero in the denominator of this inverse sum rules.
This greatly simplifies the process of poles detection since we just have to find the roots of a real function.
A more detailed analysis concerning the correspondence between finite size instabilities in finite nuclei and infinite matter will be the subject
of a forthcoming article~\cite{SchunckTODO}.

This work is organized as follows: in section~\ref{sect:linear} we summarize the
different components of the EDF and recall the main steps of the LR formalism in nuclear matter presented in article I~\cite{Davesne09}.
In particular we give explicit expressions of the RPA responses in terms of 
the coupling constants of the Skyrme functional.
We also establish the expressions of the first odd moments 
of the strength function in each channel. In section III, we present the results
concerning the detection of instabilities and for completeness we also show the resulting Landau parameters.
Further  possible developments are discussed in the conclusion.

%
\section{Linear response}
\label{sect:linear}
%

%
\subsection{Response functions and energy density functional}
%

The response functions, $\chi^{S,M,I}(\omega,\mathbf{q})$,  we are interested in are formally defined as the response of the infinite medium to external probes of the type
${\hat Q}^{{\rm (S,M,I)}} = \sum_j e^{i \bq \cdot \br_j} \, \Theta_j^{{\rm (S,M,I)}}$ where $S$ $(M)$ is the spin (its projection along the $z$-axis), $I$ the isospin and the operators $\Theta_j^{{\rm (S,M,I)}}$ are given in Table I. Following the notation of Garcia-Recio \emph{et al.}~\cite{Garcia92},  we have
\begin{eqnarray}
\label{reponse}
\chi^{{\rm (S,M,I)}}(\omega,q) = & & \frac{1}{V}\sum _n |\langle n |{\hat Q}^{{\rm (S,M,I)}}|0\rangle|^2 \\
                          & & \left(\frac{1}{\omega-E_n+i\eta}+\frac{1}{-\omega-E_n+i\eta } \right),\nonumber
\end{eqnarray}
where $\omega,\mathbf{q}$ are respectively the transferred energy and momentum, the sum is on all excited states $|n\rangle$ with energy $E_n$ and $V$ is a quantification volume (see ref.~\cite{FW71} for a detailed discussion).
Without any residual particle-hole, \emph{ph}, interaction, the above expression reduces to the usual Lindhard function. Switching  the $ph$ interaction on, the response functions can be determined with the use of the RPA formalism. Such calculations have already been published in the literature with Skyrme point-like interactions which incorporate only the central part \cite{Garcia92,Les06a} or the spin-orbit contribution \cite{Margueron06} as well. More recently, in article 
I, we generalized the previous calculations by taking into account the tensor part, which reveals to be very important quantitatively.
However, in article I we expressed  the response functions with respect to the usual coupling constants of the Skyrme effective interaction~: $\{t_i,x_i\}, i=0,..,3$, $W_0$ and $t_e,t_o$ respectively for the central, the spin-orbit and the tensor part. In the present article we write them with an energy density functional (EDF) as a starting point. This has the great advantage to be more general in the sense that all the coefficients can be now considered as independent one from each other. The parameter space is thus enlarged, allowing by instance more flexibility for the description of nuclei. In the context of forthcoming articles on instabilities, it will allow us to study precisely the role of each of these coefficients.
To be specific, we consider hereafter the following EDF (see article I for notations)

\begin{widetext}
\begin{eqnarray}
\label{eq:EF:full}
\mathcal{E}_{\text{Skyrme}}
& = & \int \! d^3r \sum_{t=0,1}
      \bigg\{  C^\rho_t [\rho_0] \, \rho_t^2
             + C^s_t    [\rho_0] \, \vec{s}_t^2
             + C^{\Delta \rho}_t  \rho_t \Delta \rho_t
             + C^{\nabla s}_t     (\nabla \cdot \vec{s}_t)^2
             + C^{\Delta s}_t     \vec{s}_t \cdot \Delta \vec{s}_t
             + C^\tau_t           ( \rho_t \tau_t - \vec{j}_t^2 )
      \nn \\
&   &
             + C^{T}_t \Big(  \vec{s}_t \cdot \vec{T}_t
                 - \sum_{\mu, \nu = x}^{z} J_{t, \mu \nu} J_{t, \mu \nu} \Big)
             + C^{F}_t \Big[  \vec{s}_t \cdot \vec{F}_t
                  - \tfrac{1}{2} \Big( \sum_{\mu = x}^{z} J_{t,\mu \mu}
                                 \Big)^2
                 - \tfrac{1}{2}
                   \sum_{\mu, \nu = x}^{z} J_{t, \mu \nu} J_{t, \nu \mu}
                \Big]
      \nn \\
&   &
             + C^{\nabla \cdot J}_t ( \rho_t \nabla \cdot \vec{J}_t
                                    + \vec{s}_t \cdot \nabla \times \vec{j}_t)
      \bigg\}
\,.
\end{eqnarray}
\end{widetext}

\noindent When the EDF is derived from a Skyrme interaction, the coupling constants can be re-expressed, following the notation of article I, in terms of $A_{t}$ and $B_{t}$ coefficients.
The coupling constant written $A_{t}$ depend on the central and spin-orbit part of the interaction ($i.e.$ $C^{\rho}_{t}=A^{\rho}_{t}$, $C^{\Delta \rho}_{t}=A^{\Delta\rho}_{t}$, $C^{\tau}_{t}=A^{\tau}_{t}$ and $C^{s}_{t}=A^{s}_{t}$, $C^{\nabla J}_{t}=A_{t}^{\nabla J}$) and  the ones written $B_{t}$ depend on the tensor part ($i.e$  $C^{\nabla s}_{t}=B^{\nabla s}_{t}$ and $C^{F}_{t}=B^{F}_{t}$), but can also contribute to the central part of the interaction ($i.e.$ $C^{T}_{t}=A^{T}_{t}+B^{T}_{t}$ and $C^{\Delta s}_{t}=A^{\Delta s}_{t}+B^{\Delta s}_{t}$).
The expressions of the coupling constants as functions of the parameters of the interaction can be found in article I.
The procedure used to obtain the residual interaction is then no longer based, as in 
article I, on the determination of the matrix elements of the particle-hole interaction from the Skyrme one, but on the direct double derivation with respect to the one-body density of the EDF. The results concerning the residual interaction coming from the tensor part are summarized in \citeAppendix{app:phme} while the response functions for infinite nuclear matter are explicitly written in \citeAppendix{app:responses}.

For completeness we also give the Landau parameters.
To obtain their expression, we have to take the limit 
$\mathbf{q}\rightarrow0$ and $\mathbf{q}_{1,2}\rightarrow \mathbf{k}_{F}$

\be
V_{ph}^{\text{Landau}}( \mathbf{k}_{F}, \mathbf{k}_{F}) = 
\lim_{\bq \rightarrow0,\bq_{1,2} \rightarrow \mathbf{k}_{F}} V_{ph}(\bq,\bq_1,\bq_2).
\ee

\noi 
According to ref. \cite{Liu,Colo10}, the most general form of the residual 
interaction in the Landau limit is

\bwt

\bqr
\label{landau}
V_{ph}^{\text{Landau}} = \delta ( \br_1 - \br_2 )  \, N_{0}^{-1} \, \sum_{\ell} \, 
\left\{ F_{\ell} + F_{\ell}' \, \hat{\tau}_{a} \circ \hat{\tau}_{b} 
                 + \left(G_{\ell}
                 + G_{\ell}' \hat{\tau}_{a} \circ \hat{\tau}_{b}\right) \vsigma_{a} \cdot \vsigma_{b}
                 + \frac{k_{12}^{2}}{k_{F}^{2}} \, H_{\ell}  \, S_{ab} 
                 + \frac{k_{12}^{2}}{k_{F}^{2}} \, H_{\ell}' \, S_{ab} \hat{\tau}_{a} \circ \hat{\tau}_{b} 
\right\} \, P_{\ell} ( \cos \theta )
\eqr

\ewt
 
\noi 
where $N_{0}^{-1} = \frac{\hbar^{2} \pi^{2}}{2 m^{*}k_{F}}$ is the usual normalization factor given 
here for the symmetric infinite nuclear matter,
$\bk_{12} = \left( \bk_1 - \bk_2 \right)$ and 
$S_{ab} = 3( \hat{\bq}_{12} \cdot \vsigma_a)( \hat{\bq}_{12} \cdot \vsigma_b )- \vsigma_a \cdot \vsigma_b$ 
, where the symbol $\hat{\bq}_{12}$ indicates a vector of unitary length. 
One can express the product of momentum and Pauli matrices  as
$(\hat{\bk}_{12} \cdot \vsigma_a )(\hat{\bk}_{12} \cdot \vsigma_b) 
= \frac{1}{3} S_{ab} + \frac{1}{3} \vsigma_a \cdot \vsigma_b$.
It is important to notice that the $H$ coefficients are functions of $\cos\theta$ and are only related to the tensor part of the interaction \cite{Friman,Colo10,Backman78,Dabrowski76,Brown77}.
\noi 
In our case the residual interaction in the Landau limit reads

\bwt

\bqr
\label{limitL}
V_{ph}^{\text{Landau}}(\mathbf{k}_{F},\mathbf{k}_{F}) & = & \frac{1}{4} W^{(0,0)}_{1,L}
                      +  \frac{1}{4}W^{(0,1)}_{1,L} \hat{\tau}_{a} \circ \hat{\tau}_{b}
                     +  \frac{1}{4} W^{(1,0)}_{1,L} \text{\boldmath$\sigma$}_{a} \cdot \text{\boldmath$\sigma$}_{b} 
                     + \frac{1}{4} W^{(1,1)}_{1,L} \text{\boldmath$\sigma$}_{a} \cdot \text{\boldmath$\sigma$}_{b}                                                       \, \hat{\tau}_{a} \circ \hat{\tau}_{b}                                       \\
                       & + & \frac{1}{4} \left[ W^{(0,0)}_{2,L}
                      + W^{(0,1)}_{2,L} \hat{\tau}_{a} \circ \hat{\tau}_{b}
                     +   W^{(1,0)}_{2,L} \text{\boldmath$\sigma$}_{a} \cdot \text{\boldmath$\sigma$}_{b} 
                     +  W^{(1,1)}_{2,L} \text{\boldmath$\sigma$}_{a} \cdot \text{\boldmath$\sigma$}_{b} 
                                      \, \right] \, \left[ 2k_{F}^{2}-2k_{F}^{2}\cos\theta \right]             \nn \\
                       & + & \frac{1}{3} \, \left[ \, k_{F}^{2}C_{0}^{F} \, + \, k_{F}^{2}C_{1}^{F} \, \hat{\tau}_{a} \circ \hat{\tau}_{b} 
                                         \, \right] \,  \frac{k_{12}^{2}}{k_{F}^{2}} \, S_{ab},        \nn \\
\nn                                         
\eqr

\ewt

\noi 

where the $W_{j,L}^{(S,I)}$ with $j=1,2$ coefficients  are  given in~\citeAppendix{app:wcoef}.
We checked that in the case of a Skyrme force we get the same values as in ref. \cite{Colo10}.

%
\subsection{Sum rules and moments of the strength function}
%

It should be noted that the quantity of interest is not directly the response function itself discussed in the previous paragraph, but merely
  $S^{(\alp)}(\mathbf{q},\omega)$, usually called dynamical structure function,  which is, 
at zero temperature, proportional to the imaginary part 
of the response function at positive energies
\be 
 S^{(\alp)}(\mathbf{q},\omega)
   = -\frac{1}{\pi} \, \rm{Im} \, \chi^{(\alp)}(\mathbf{q},\omega)\,. 
\label{eq:sfunc}
\ee

\noindent On the other side, the k-moments, which are defined as moments per particle in infinite matter, read
\be 
M^{(\alp)}_k(\mathbf{q})=\sum _n E_n^k|\langle n |{\hat Q}^{{\rm (\alp)}}|0\rangle|^2. 
\ee

\noindent  After some manipulation we can express them  as an integral of the dynamical structure function $S^{(\alp)}(\mathbf{q},\omega)$ shown in \citeqdot{eq:sfunc} as

\be 
M^{(\alp)}_k(\mathbf{q}) \, = \, \int_0^{\infty} \, d \omega \, \omega^k \, S^{(\alp)}(\mathbf{q},\omega) \q .
\label{eq:mk}
\ee

\noindent Moreover, because of its intrinsic analytic properties, the response function satisfies a dispersion relation. As a consequence, moments can be obtained analytically through appropriate expansions in power series of $\omega$ \cite{Garcia92}

\bi
   \item for $\omega \rightarrow +\infty$, the positive odd order
         moments read
         
\be
\label{relM1M3}
\chi^{(\alpha)}(\omega,\mathbf{q}) \approx \phantom{-}2 \rho
                                        \sum_{p=0}^{+\infty} \, (\omega)^{-(2p+2)} M^{(\alpha)}_{2p+1}(\mathbf{q})  ,
\ee

\noi
and can be used for the calculation of the $M_{1}$ Energy Weighted Sum Rule (EWSR) 
and the $M_{3}$ Cubic Energy Weighted Sum Rule (CEWSR).

   \item for $\omega \rightarrow 0$, the negative odd order moments can be extracted as

\be
\label{relMm1}
\chi^{(\alpha)}(\omega,\mathbf{q}) \approx -2 \rho
                                       \sum_{p=0}^{+\infty} \, (\omega)^{2p} M^{(\alpha)}_{-(2p+1)}(\mathbf{q})   , 
\ee

\noi
which will be used for the $M_{-1}$ Inverse Energy Weighted Sum Rule (IEWSR).

\ei

\noindent In the above formula $\rho$ represents the  density of the system.

The situation is the following~: we have at our disposal two expressions for $M^{(\alp)}_k$~: the first one (Eq. (\ref{eq:mk})) is purely numerical and implies the whole response function; the second one Eq. (\ref{relM1M3}) (or Eq. (\ref{relMm1})), which originates from the dispersion relation satisfied by the response function is an analytic expansion  (see the next paragraph for explicit expression). Both should  coincide with very high accuracy. When it is not the case, this means that the dispersion relation is no longer valid or, in other words, that a pole occurs. Thus, a discrepancy between the different expressions for the sum rules will be interpreted as the presence of a pole.

We shall now enter the details and discuss separately the three important sum rules $M_1, M_3, M_{-1}$. Since the other cases can be obtained by switching off the appropriate coupling constants, the general case with the tensor will be considered only. 
As stated previously, all the expressions given below for these
sum rules are valid for a general Skyrme EDF given in Eq.(\ref{eq:EF:full}) in which all the coupling constants could be considered as independent of each other.

%
\subsubsection{Energy weighted sum rule}

Making the appropriate asymptotic expansion of the response functions written in \citeAppendix{app:responses}, we obtain for each channel

\bqr\label{eq:m1_0I}
M^{{\rm (0,I)}}_{1}        =  \frac{q^2}{2m^*}\left[ 1 -  \frac{m^* \rho}{2} W_{2}^{(0,\text{I})}\right] ,  
\eqr

\bqr\label{eq:m1_10I} 
M^{{\rm (1,0,I)}}_{1} = \frac{q^2}{2m^*} \left[1 -  \frac{m^* \rho}{2} \left(W_{2}^{(1,\text{I})}+4B_{\text{I}}^T+4B_{\text{I}}^F\right) \right], 
\eqr         

\bqr \label{eq:m1_11I}                                    
M^{{\rm (1,\pm 1,I)}}_{1}  =  \frac{q^2}{2m^*}\left[ 1 - \frac{m^* \rho}{2} \left(W_{2}^{(1,\text{I})}+4B_{\text{I}}^T\right) \right].  
\eqr

\noindent If one now takes into account the expression of the isoscalar effective mass, 
\emph{i.e.} $m/m^* = 1 + 2 m \rho C_0^{\tau}$, one can rewrite the expressions of the EWSR in terms of the coupling constants of the Skyrme EDF as

\be
M^{{\rm (S,M,I)}}_{1}  \, = \, \frac{q^2}{2m} 
                         \, + \, q^2 \, \rho \, \alpha^{{\rm (S,M,I)}}  ,  \\
\label{eq:m1alpha}
\ee

\noi
where $\alpha^{{\rm (S,M,I)}}$ is a sum of contributions corresponding 
to each part of the Skyrme EDF which is considered, central and 
tensor parts of the Table I for instance.
The free part $\frac{q^2}{2m}$ corresponds to the kinetic part of the 
hamiltonian since only the gradient terms of the interaction 
contribute to the corresponding $\alpha^{{\rm (S,M,I)}}$ coefficient.
Note that the spin-orbit part of the Skyrme interaction does not 
contribute to the EWSR.

In self-consistent RPA calculations, \emph{i.e.} when the same effective interaction generates 
the HF mean field and also produces the residual interaction, positive odd-order 
RPA sum rules can be calculated through the Thouless theorem, by taking 
the expectation values of appropriate operators on the HF ground state 
(see for example Bohigas \etal~\cite{Bohigas} and Lipparini \etal~\cite{Lipparini}
for the details of this technique).
For the $M^{{\rm (S,M,I)}}_{1}$ EWSR one can write

\be
\label{eq:m1comm}
M^{{\rm (S,M,I)}}_{1}  \, = \, \pdemi \,
\la 0 \vert \left[ {\hat Q}^{{\rm (S,M,I)}} , \left[ {\hat Q}^{{\rm (S,M,I)}} , H \right] \right] \vert 0 \ra,
\ee

\noi
calculated for each $(\alp)$ channel with the operator 
${\hat Q}^{{\rm (S,M,I)}}=\sum_j e^{i \bq \cdot \br_j} \, \Theta_j^{{\rm (S,M,I)}}$ given 
in Table I and with the hamiltonian $H$ built up with the zero range Skyrme effective interaction.
We have checked, after some tedious calculations, that this result coincides exactly, as it should be, with that obtained with the asymptotic expansion of the response. Note that the double commutator technique (Eq. \ref{eq:m1comm})
uses the full hamiltonian $H$ of the system with a Skyrme interaction and it can not be used with a generalized EDF which does not derive from a Skyrme interaction.

%
\begin{table}[h!]
\label{tabm1alpha}
\caption{Operators used in each ${\rm (S,M,I)}$ channel. 
Columns 3 and 4 give the central and tensor contributions to the EWSR respectively.
$\sigma_i^0, \sigma_i^{\pm 1}$ 
($\tau_i$ respectively) are the standard components of the vector $\vsigma_i$ defined 
as $\sigma_i^0 = \sigma_i^z$ and 
$\sigma_i^{\pm 1} = \mp \left( \sigma_i^x \pm i \sigma_i^y \right) / \sqrt{2}$.}
\begin{ruledtabular}
\begin{tabular}{cccccccccc}
 ${\rm (S,M,I)}$ &&& ${\Theta}_i^{{\rm (S,M,I)}}$     &&& $\alpha^{{\rm (S,M,I)}}$           
                                                    &&& $\alpha^{{\rm (S,M,I)}}$                   \\ 
                 &&&                                &&& $({\rm central})$           
                                                    &&& $({\rm tensor})$                           \\
\noalign{\smallskip}\hline\noalign{\smallskip} 
 $(0,0,0)$       &&& 1                              &&& 0                                         
                                                    &&& 0                                          \\
 $(0,0,1)$       &&& $\tau_i^0$                     &&& $A_0^{\tau} - A_1^{\tau}$ 
                                                    &&& 0                                          \\
 $(1,0,0)$       &&& $\sigma_i^0$                   &&& $A_0^{\tau} - A_0^T$
                                                    &&& $- B_0^T - B_0^F$                          \\
 $(1,\pm 1,0)$   &&& $\sigma_i^{\pm 1}$             &&& $A_0^{\tau} - A_0^T$
                                                    &&& $- B_0^T$                                  \\
 $(1,0,1)$       &&& $\sigma_i^0 \, \tau_i^0$       &&& $A_0^{\tau} - A_1^T$     
                                                    &&& $- B_1^T - B_1^F$                          \\
 $(1,\pm 1,1)$   &&& $\sigma_i^{\pm 1} \, \tau_i^0$ &&& $A_0^{\tau} - A_1^T$
                                                    &&& $- B_1^T$                                  \\
\end{tabular}
\end{ruledtabular}
\end{table}
%

\subsubsection{Cubic energy weighted sum rule}

Making the expansions of the responses (see \citeAppendix{app:responses}), we obtain successively for each channel

\bwt

\bqr
M^{{\rm (0,I)}}_{3}  & = & q^4 \left( \frac{k_F^2}{2m^{*3}} \right)   \, 
                           \left[ \pdemi m^* \rho  \wdzI   - 1 \right]^2  \, \\ 
                     & \times & \left\{ \tfrac{3}{5} + k^2 + \pdemi k^2 m^* \rho  \wdzI 
                           + \pdemi \mktpi \left( \wuzI + 2k_F^2 \wdzI \right) \right\}       \nn    \, , \\
\nn
\label{eq:m3_0I}
\eqr

\bqr
M^{{\rm (1,0,I)}}_{3} & = & q^4 \, \left( \frac{k_F^2}{5m^*} \right) \left[ \rho B_I^F \right]^2   
                            \, \left[  m^{*}\rho\left(W_{2}^{(1,\text{I})}+4B_{\text{I}}^{T}+4B_{\text{I}}^{F}\right) - 1 \right]                             \\
                        & + & q^4 \left( \frac{k_F^2}{2m^{*3}} \right) 
                            \, \left[ \pdemi m^{*}\rho \left(W_{2}^{(1,\text{I})}+4B_{\text{I}}^{T}+4B_{\text{I}}^{F}\right)- 1 \right]^2                                \nn \\
                   & \times & \left\{ \tfrac{3}{5} + k^2 + \tfrac{6}{5} m^* \rho B_{\text{I}}^F + \pdemi m^{*}\rho k^2 \left( W_{2}^{(1,\text{I})}+4B_{\text{I}}^{T}\right)  
                          + \pdemi \mktpi 
                            \left[\tilde{W}_1^{{\rm (1,0,I)}}   + 2k_F^2\left(W_{2}^{(1,\text{I})}+4B_{\text{I}}^{T} \right) \right] \right\}     , \nn \\
\nn\\
\tilde{W}_1^{{\rm (1,0,I)}}&=&W_1^{{\rm (1,I)}}+8q^{2}\left(B_{\text{I}}^{\nabla s}-B_{\text{I}}^{\Delta s}\right)-2q^{2}B^{T}_{\text{I}},\nn 
\nn
\label{eq:m3_10I}
\eqr

\bqr
M^{{\rm (1,\pm1,I)}}_{3} & = & q^4 \left( \frac{k_F^2}{10m^*} \right) 
                               \left[ \rho B_I^F \right]^2 \left[ m^{*}\rho\left( W_{2}^{(1,\text{I})}+4B_{\text{I}}^{T} \right)- 1 \right]                         \\
                           & + & q^4 \left( \frac{k_F^2}{2 m^{*3}} \right) \left[ \pdemi m^{*}\rho \left( W_{2}^{(1,\text{I})}+4B_{\text{I}}^{T} \right)- 1 \right]^2  \nn \\
                      & \times &     \left\{ \tfrac{3}{5} + k^2 + \tfrac{2}{5} m^* \rho B_I^F + \pdemi k^2 m^{*}\rho \left( W_{2}^{(1,\text{I})}+4B_{\text{I}}^{T} \right)
                             + \pdemi \mktpi              
                               \left[ \tilde{W}_1^{{\rm (1,\pm 1,I)}} + 2k_F^2 \left( W_{2}^{(1,\text{I})}+4B_{\text{I}}^{T} \right) \right] \right\}           \, , \nn \\
\nn\\
\tilde{W}_1^{{\rm (1,\pm 1,I)}}&=&W_1^{{\rm (1,I)}}-\left( 8B^{\Delta s}_{\text{I}}+2B^{T}_{\text{I}}\right)q^{2}, \nn 
\label{eq:m3_11I}
\eqr

\ewt

\noi
with the $W_{i}^{\text{(S,I)}}$ coefficients given in~\citeAppendix{app:wcoef} and the usual relations~:
$\rho = 2 k_F^3/(3\pi^2), k = \frac{q}{2k_F}$.                              

In principle the $M^{{\rm (S,M,I)}}_{3}$ CEWSR can be also obtained analytically
from the commutator machinery briefly described in the previous paragraph but the operator to be 
considered involves now a triple commutator, \emph{i.e.} three times the interaction.
The calculation becomes then very tedious and has been checked  
only for the central part of the Skyrme interaction~\cite{Lallouet11}
which gives a CEWSR which does not depend on the value of  spin projection~$M$ 

\bwt

\bqr
M^{{\rm (S,M,I)}}_{3} & = & q^4 \left( \frac{k_F^2}{2 m^{*3}} \right) \left[ \pdemi m^{*}\rho  W_{2}^{\text{(S,I)}} - 1 \right]^2  \nn \\
                      & \times &     \left\{ \tfrac{3}{5} + k^2  + \pdemi k^2 m^{*}\rho W_{2}^{(\text{S,I})}
                             + \pdemi \mktpi              
                               \left[W_1^{{\rm (S,I)}} + 2k_F^2  W_{2}^{(\text{S,I})}  \right] \right\}.   \\
\nn
\eqr

\ewt

\subsubsection{Inverse energy weighted sum rule}

This moment cannot be obtained through the commutator machinery but only using
appropriate constrained Hartree-Fock calculations through the well-known
dielectric theorem~\cite{Bohigas}.

Thus, we use again the appropriate expansion of the expressions given 
in~\citeAppendix{app:responses} to obtain finally

\bwt

\bqr
M_{-1}^{{\rm (0,I)}} & = & f(k) \, \left( \frac{3m^*}{2k_F^2} \right) 
\, \left\{ - 48 \left[ m^* \rho k C_I^{\nabla J} \right]^2 \,
   \frac{f(k) \coka}{8 - m^* \rho \coka \left[ W_{2}^{(1,\text{I})}+4B^{T}_{\text{I}} - 2B_I^F \right]}         \right. \nn \\
                       & - & \tfrac{3}{64} \, 
                       \left[ m^* \rho f(k) \left( 1 - k^2 \right) \wdzI \right]^2         
                     + \left[ 1 + \tfrac{3}{8} m^* \rho \wdzI \right]^2                         \nn \\ 
                       & + & \left. f(k) \left[ \left( \mkdpi \right) \wuzI
                                 + \tfrac{3}{4} m^* \rho \left( 1 - k^2 \right) \wdzI
                                 - \tfrac{1}{32} \left( 3 + 13 k^2 \right) 
                                                 \left( m^* \rho \wdzI \right)^2  \right] \, \right\}^{-1} , \\
\nn
\label{eq:m-1_0I}
\eqr

\bqr
M_{-1}^{{\rm (1,0,I)}} & = & f(k) \left( \frac{3m^*}{2k_F^2} \right) 
\, \left\{ \left[ 1 + \tfrac{1}{8} m^* \rho \left( 3 W_{2}^{\text{1,I}}+12B^{T}_{\text{I}} + 8B_I^F \right) \right]^2             \right. \nn \\
                         & - & \tfrac{3}{64} \,
                   \left[ f(k) \left( k^2-1 \right) m^* \rho \right]^2 \left[ W_{2}^{\text{1,I}}+4B^{T}_{\text{I}}\right]^2             \nn \\
                         & + & f(k) \, \left[ \left( \mkdpi \right) \tilde{W}_{1}^{{\rm (1,0,I)}} 
                                        + \tfrac{3}{4} m^* \rho ( 1 - k^2 ) \left( W_{2}^{\text{1,I}}+4B^{T}_{\text{I}}\right)
                                        - \tfrac{3}{2} k^2 m^* \rho \left( 4 B_I^F \right) \right. \nn \\
                         & - & \tfrac{1}{32} m^{*2} \rho^2 \left. \left. \, 
                        \left( 96 \left(1+  k^2\right) [B_I^F]^2 
                             + 24 \left(1+ 3k^2\right)  B_I^F  \left( W_{2}^{\text{1,I}}+4B^{T}_{\text{I}}\right)  
                             +    \left(3+13k^2\right) \left( W_{2}^{\text{1,I}}+4B^{T}_{\text{I}}\right)^2 \right) \right]
\, \right\}^{-1}, \\
\nn          
\label{eq:m-1_10I}                                                      
\eqr

\bqr
M_{-1}^{{\rm (1,\pm1,I)}}   & = & f(k) \left( \frac{3m^*}{2k_F^2} \right) 
\, \left\{ - 24 \left[ k m^* \rho C_I^{\nabla J} \right]^2 \, 
                \frac{f(k) \coka}{8 - m^* \rho \coka \wdzI }        \right.  \nn \\
                     & + & \left[ 1 + \tfrac{3}{8} m^* \rho 
                           \left(  W_{2}^{\text{1,I}}+4B^{T}_{\text{I}} + \tfrac{2}{3} B_I^F \right) \right]^2\nn\\
                       &  -& \tfrac{3}{64} \left[ m^* \rho f(k) \left( 1 - k^2 \right) \right]^2 
                           \left[ 20 [ B_I^F ]^2 + 4 B_I^F \left( W_{2}^{\text{1,I}}+4B^{T}_{\text{I}}\right) +  \left( W_{2}^{\text{1,I}}+4B^{T}_{\text{I}}\right)^2 \right]             \nn \\
                     & + & f(k) \left[ 
                           \left( \mkdpi \right) \tilde{W}^{(1,\pm1,I)}_{1} 
                                + \tfrac{3}{4} m^* \rho ( 1 - k^2 ) 
                                  \left( 2B_I^F +  W_{2}^{\text{1,I}}+4B^{T}_{\text{I}} \right)                       \right. \nn \\                
                     & - & \left.\left. \tfrac{1}{32} m^{*2} \rho^2 
                           \left( 4[B_I^F]^2 \left( 9 - k^2 \right) + 16 k^2 B_I^F  \left( W_{2}^{\text{1,I}}+4B^{T}_{\text{I}}\right)
                                                     + \left(3+13k^2\right) \left( W_{2}^{\text{1,I}}+4B^{T}_{\text{I}}\right)^2 \right)
                                \right]            
\, \right\}^{-1}    , \\           
\nn
\label{eq:m-1_11I}
\eqr

\ewt
with $f(k) = \frac{1}{2} \, \left[ 1 + \frac{1}{2k} \left( 1 - k^2 \right) 
                              \log \left( \frac{k+1}{k-1} \right) \right]$.
Since the instabilities we are looking for, are related with poles of the response functions at zero energies, 
this sum rule will be shown in the next part to be of fundamental importance for the detection and therefore prediction of instabilities. Since in the definition of $M_{-1}$, the contribution of the low-energy part is more important because of the factor $1/\omega$ in the integrand, this sum rule is more sensitive to the poles than the others.

\section{Results}
\subsection{Response functions}

We have already discussed in article I the fact that the tensor may contribute significantly to the response functions. Here, we precise several related aspects in view of the forthcoming discussion about instabilities.

Quite generally, in $S=0$ channel, the tensor terms do not affect qualitatively the response;
all tests performed using TIJ tensor interactions discussed in~\citerefdot{Lesinski07} exhibit the same
qualitative behavior. The situation is quite different in $S=1$ channels; the effect from the tensor
terms is large whatever the value of the spin projection $M$ is.
Actually, depending on the values of the transferred momentum
$q$ and the density $\rho$,  the response functions can even increase significantly
and diverge for finite $q$ for a certain critical
density $\rho_c$. As illustrated in Fig. 1, one can typically observe two types of extremes phenomena~: the first one (left panel) corresponds to an accumulation of strength at finite energy (and low transfer momentum) and is related to the zero sound mode whereas the second one (right panel) is associated to a pole at zero energy (and finite momentum).
Although a one-to-one correspondence between infinite matter and nuclei is obviously not trivial, preliminary tests seem to show that 
the latter divergence actually reveals the presence of instabilities observed in nuclei \cite{Les06a},
with the appearance of domains with typical size of the order of $2\pi/q$ \cite{Karim11}.
The center of a nucleus effectively explores, because of fluctuations, not only the saturation
density but also some larger values for which one may observe a divergence
of the response functions. In the following, we will concentrate ourselves on the detection of such poles.

\begin{figure*}[htbp]
       \includegraphics[clip,scale=0.3,angle=-90]{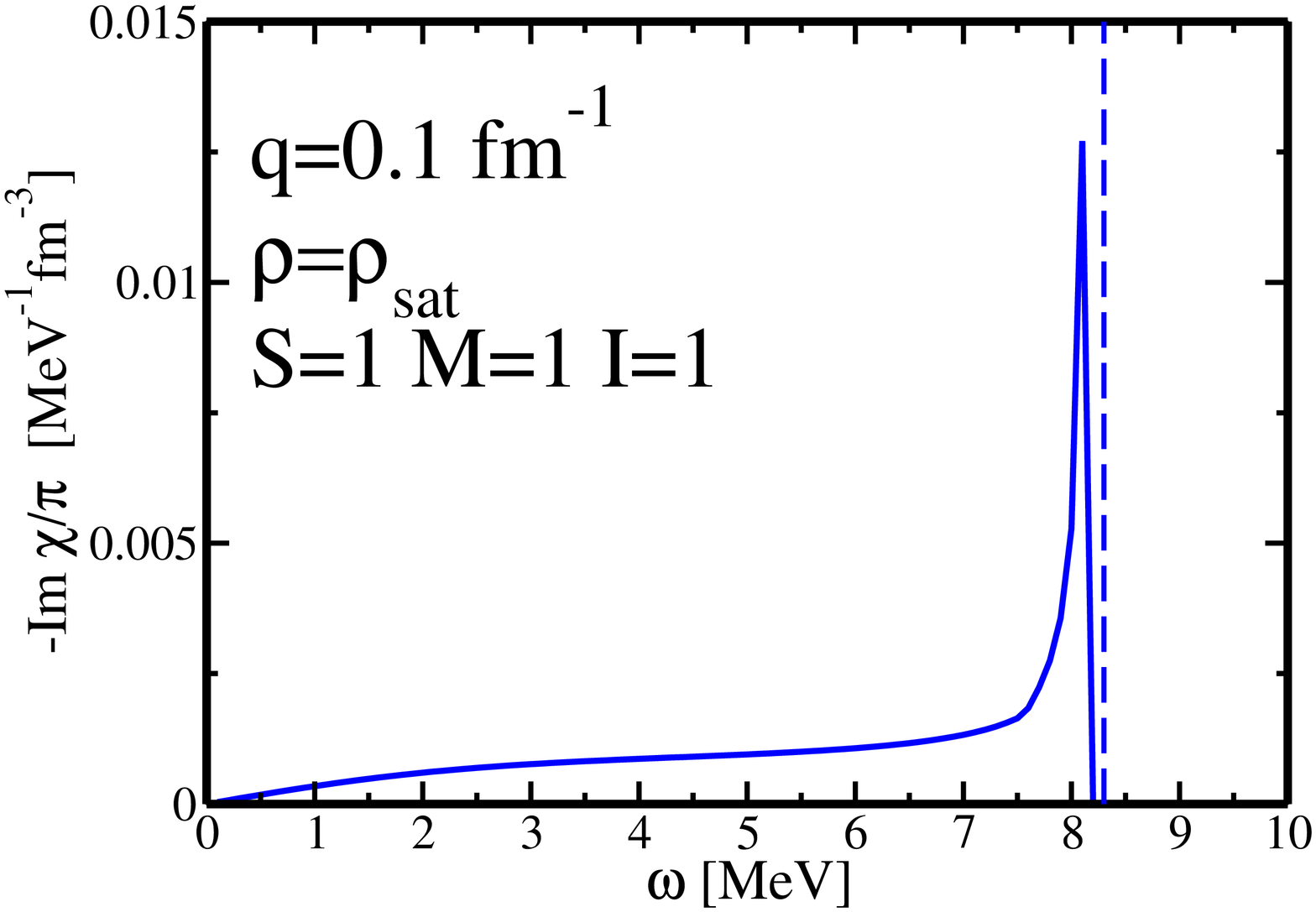}
      \includegraphics[clip,scale=0.3,angle=-90]{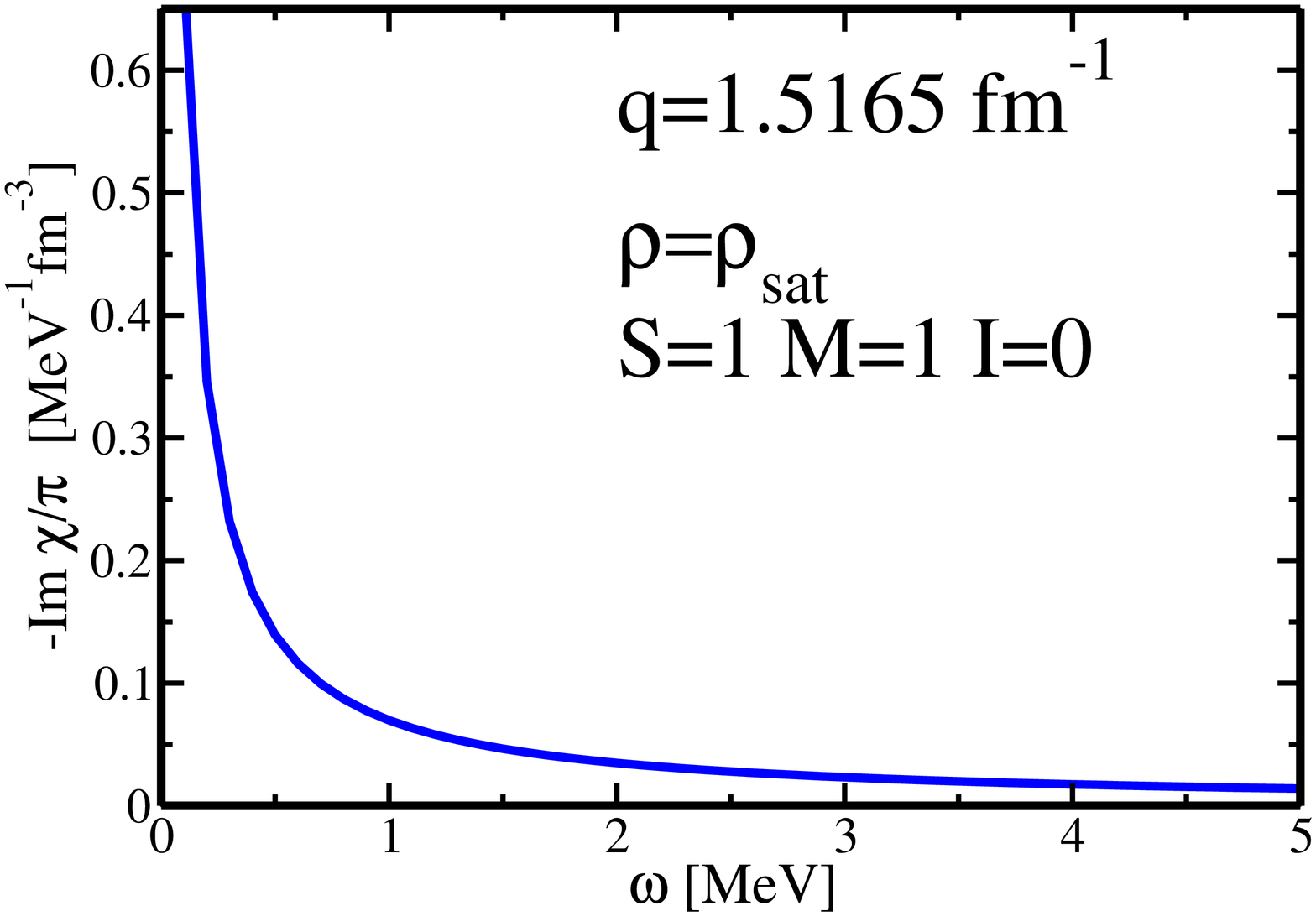}
      \caption{Nuclear response  function for two given values 
               of the transferred momentum $q$ (in fm$^{-1}$) for the 
               T44 tensor parameterization.  On the left we show the  response function in the channel $(1,1,1)$ with the discrete $ph$ transition corresponding to the zero sound. On the right we show the  response function in the channel $(1,1,0)$ in proximity of a pole.$\rho_{sat}$ is the saturation density of the system. }
\label{fig:T44_response}
\end{figure*}

\subsection{Sum rules}
\subsubsection{EWSR}

As an example, \citeFigure{fig:m1_T44} shows the EWSR 
calculated for the equilibrium density, for the $T44$ 
tensor parameterization and for all the six ${\rm (S,M,I)}$ channels.
In each case the result obtained with the integral (\citeqdot{eq:mk}) 
is compared to the exact calculation 
(\citeqssdot{eq:m1_0I}{eq:m1_10I}{eq:m1_11I}).
As expected, both results coincide, satisfying then
the sum rule.
It remains an exception for the two ${\rm (1,0,1)}$ and ${\rm (1,1,0)}$
channels where the integral calculation violates the sum rule.
This is actually due to the presence of a pole  (indicated by an arrow on Fig. 2) in the strength function at
$q \simeq 1.5$ fm$^{-1}$ for the ${\rm (1,1,0)}$ channel (see caption of Fig. 1 right panel) and at
$q \simeq 2.2$ fm$^{-1}$ for the ${\rm (1,0,1)}$ channel.
These poles which are clearly exhibited in the IEWSR (see below)
make the sum rules unphysical from and above the $q$ value of the
pole.

\begin{figure}[h]
      {\includegraphics[clip,scale=0.35,angle=-90]{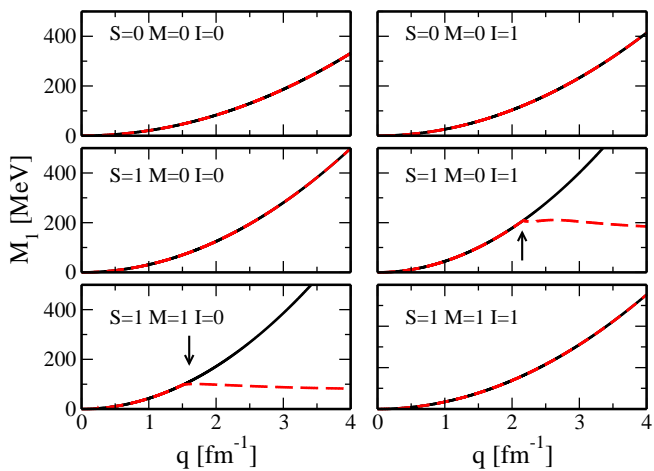}
      \caption{(Color on line) EWSR (in MeV) as a function of the 
               transferred momentum $q$ (in fm$^{-1}$) for the T44 
               tensor parameterization.
               Red dashed (black solid) line correspond to the
               integral (\citeqdot{eq:mk}) (analytical expressions,
               \citeqssdot{eq:m1_0I}{eq:m1_10I}{eq:m1_11I} respectively).
               Results are shown for the saturation density and for 
               each ${\rm (S,M,I)}$ channel.}
\label{fig:m1_T44}}
\end{figure}

\subsubsection{CEWSR}

For the same example and for the same conditions, 
\citeFigure{fig:m3_T44} shows the CEWSR.
As for the EWSR we observe a perfect correspondence between the
two calculations of the sum rule, integral or analytical expression
(see \citeqssdot{eq:m3_0I}{eq:m3_10I}{eq:m3_11I})
except in the channels which exhibit a pole in the strength function.
The same remarks as for the EWSR can be done. 
Due to the cubic energy weighted in this sum rule, the violation 
of the concerned sum rules does not appear very clearly on the figures.   

\begin{figure}[h]
      {\includegraphics[clip,scale=0.35,angle=-90]{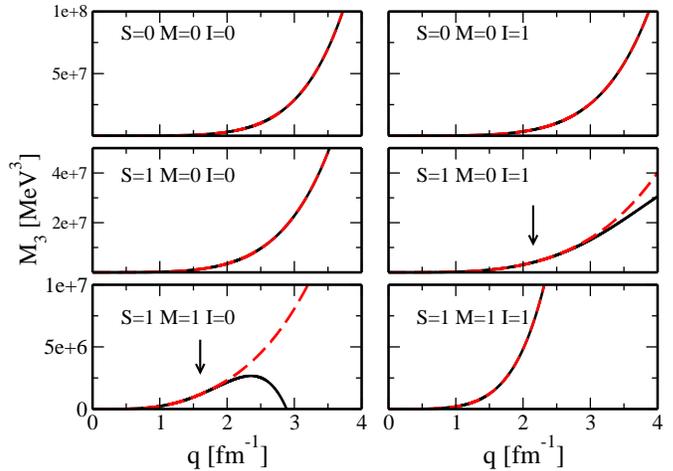}
      \caption{(Color on line) CEWSR (in MeV$^3$) as function of the 
               transferred momentum $q$ (in fm$^{-1}$) for the T44 
               tensor parameterization.
               See \citeFigure{fig:m1_T44} for other details.}
\label{fig:m3_T44}}
\end{figure}

\subsubsection{IEWSR}

Always for the same example and for the same conditions, 
\citeFigure{fig:m-1_T44} shows the IEWSR.
As for the EWSR we observe a good correspondence between the
integral and the analytical expression 
(see \citeqssdot{eq:m-1_0I}{eq:m-1_10I}{eq:m-1_11I})
except when there is a pole in the strength function.
Moreover the discrepancy in that case is very sharp so that the signature of the pole is
very clear.
This is due to the fact that for IEWSR this pole is also present
on the denominator of the analytical expressions
(see \citeqssdot{eq:m-1_0I}{eq:m-1_10I}{eq:m-1_11I}).
Thus, we show here that there is unique correspondence with the pole
observed in the strength function and the pole of the IEWSR.
An immediate consequence is that \citeqssdot{eq:m-1_0I}{eq:m-1_10I}{eq:m-1_11I} can be used in a fit protocol in order to test directly the occurrence of instabilities.

\begin{figure}[h]
      {\includegraphics[clip,scale=0.35,angle=-90]{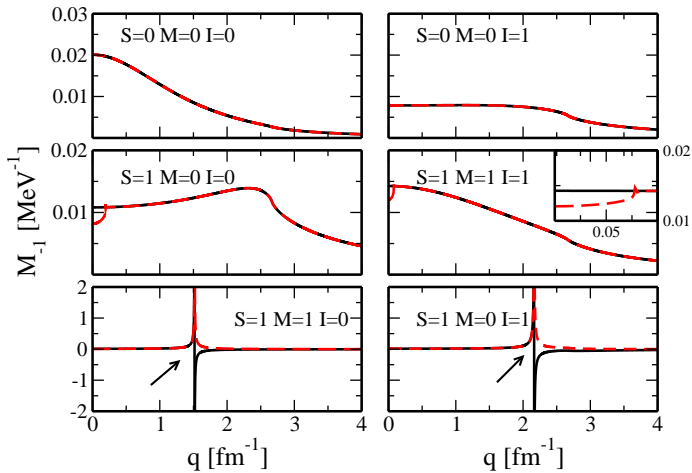}
      \caption{(Color on line) IEWSR (in MeV$^{-1}$) as function of the 
               transferred momentum $q$ (in fm$^{-1}$) for the T44 
               tensor parameterization. 
               See \citeFigure{fig:m1_T44} for other details.}
\label{fig:m-1_T44}}
\end{figure}

It should also be noticed (see the inset of Fig.\ref{fig:m-1_T44}) that a small amount of strength is missing at low $q$ in channels $(1,0,0)$ and $(1,1,1)$.
This again corresponds to the zero sound mode already shown in left panel of Fig. \ref{fig:T44_response}. and will not be discussed here.

\subsection{Instabilities}
\label{sect:instabilities}

When the response function exhibits a pole at zero energy in a given channel one can suspect 
that an unphysical instability will occur in finite nuclei if the corresponding critical density is closed to the saturation density.
The goal of this part is thus to show for typical Skyrme parametrisations whether they lead to such problem, that is the appearance of an unphysical instability when the critical density $\rho_{c}$ calculated using the IEWSR is close to the saturation density . Since we have shown that there is a direct connection between the pole 
(when it does exist) observed in the response function and the pole observed
in the $M_{-1}$ sum rule. It is easy to plot the critical densities $\rho_c$ as a function of $q$ by simply solving $1/M_{-1} (\rho_c,q) =0$ in each channel. For example, \citeFigure{fig:critical_T44} shows the behavior of the critical 
density for each ${\rm (S,M,I)}$ channel for the interaction T44. As a guide-eye, the saturation density $\rho_{sat}$ is also plotted. 
As claimed, one can clearly see that one exactly obtains the same results if one
considers the pole of the $M_{-1}$ sum rule (open circles)
or the pole of the corresponding RPA responses (dashed lines). 
For this particular parametrisation, instabilities
appear both in the ${\rm (1,0,0)}$ and ${\rm (1,1,0)}$ channels at $\rho_{c}=\rho_{sat}$.
For the (0,0) channel, we can also see on \citeFigure{fig:critical_T44} the well-known
spinodal instability.
This spinodal instability is viewed here as a two branches curve corresponding
to the two critical densities observed in a standard plot of this
spinodal curve. These two branches meet at the critical point.
This fact is due to the $C^{\Delta \rho}_{0}$ coupling constant
and it can be viewed as a surface effect.
Without this term in the functional the two branches of the
spinodal curve would turn into two parallel lines~\cite{Camille11,Ducoin07,Ducoin08a,Ducoin08b}.

\begin{figure}[htbp]
      {\includegraphics[clip,scale=0.3,angle=-90]{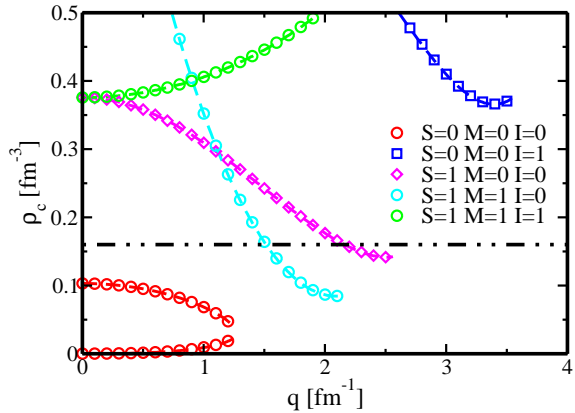}
      \caption{(Color on line) Critical densities (in fm$^{-3}$) as 
               functions of the transferred momentum $q$ (in fm$^{-1}$) 
               for the T44 tensor parameterization. 
               Open circles show the critical densities extracted from 
               the pole of the $M_{-1}$ moment while 
               the dashed lines correspond to the pole of the 
               corresponding strength function.}
\label{fig:critical_T44}}
\end{figure}

\noindent \citeFigure{fig:critical_Skyrme} displays the critical densities for some
usual Skyrme EDF.
All the Skyrme EDF exhibit the same physical spinodal instability but the behaviors
of the critical densities in the other channels are very different
and depend strongly of the parameterization under consideration.

\begin{figure*}
\bc
      \includegraphics[clip,scale=0.3,angle=-90]{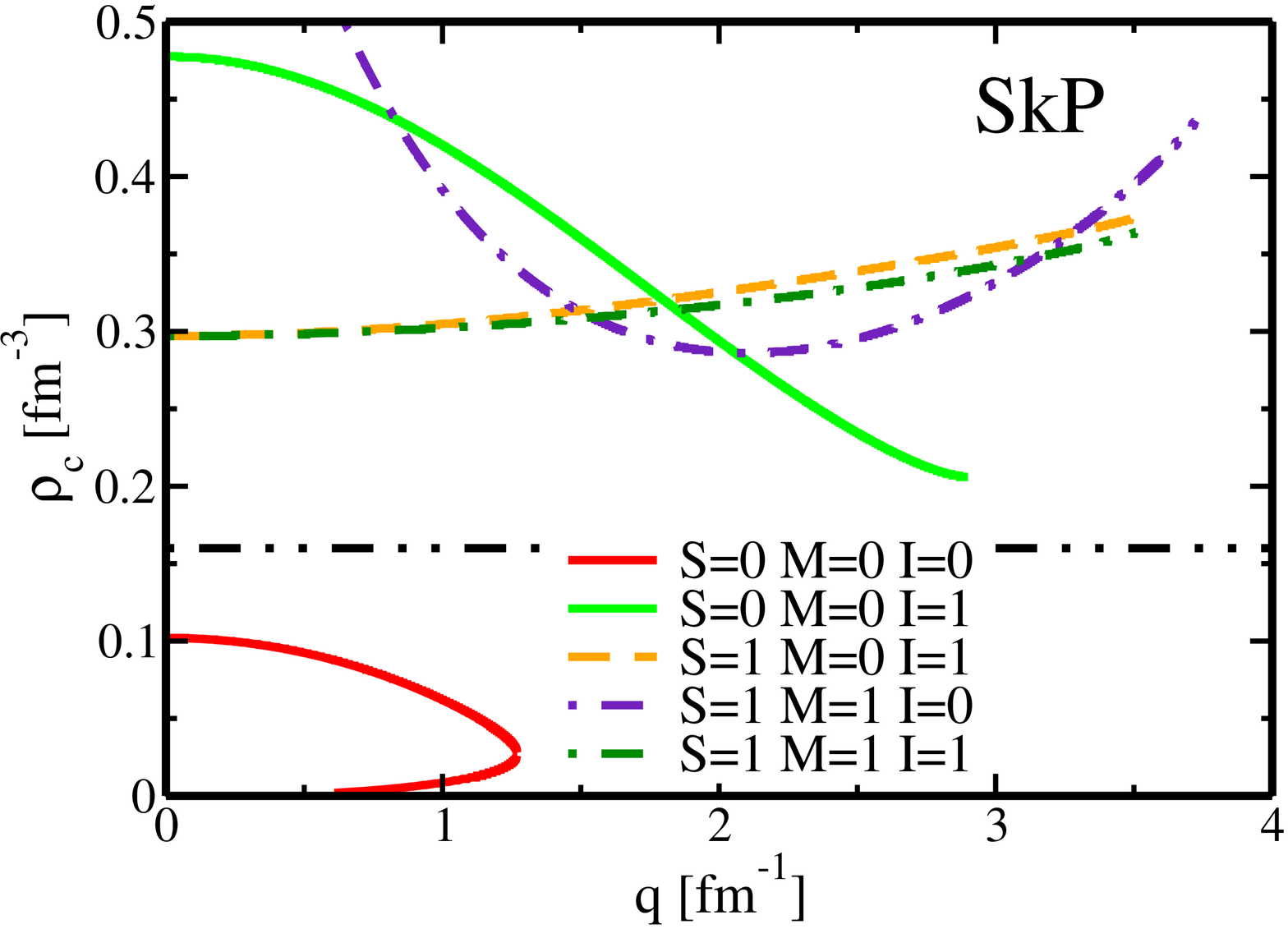}   
      \includegraphics[clip,scale=0.3,angle=-90]{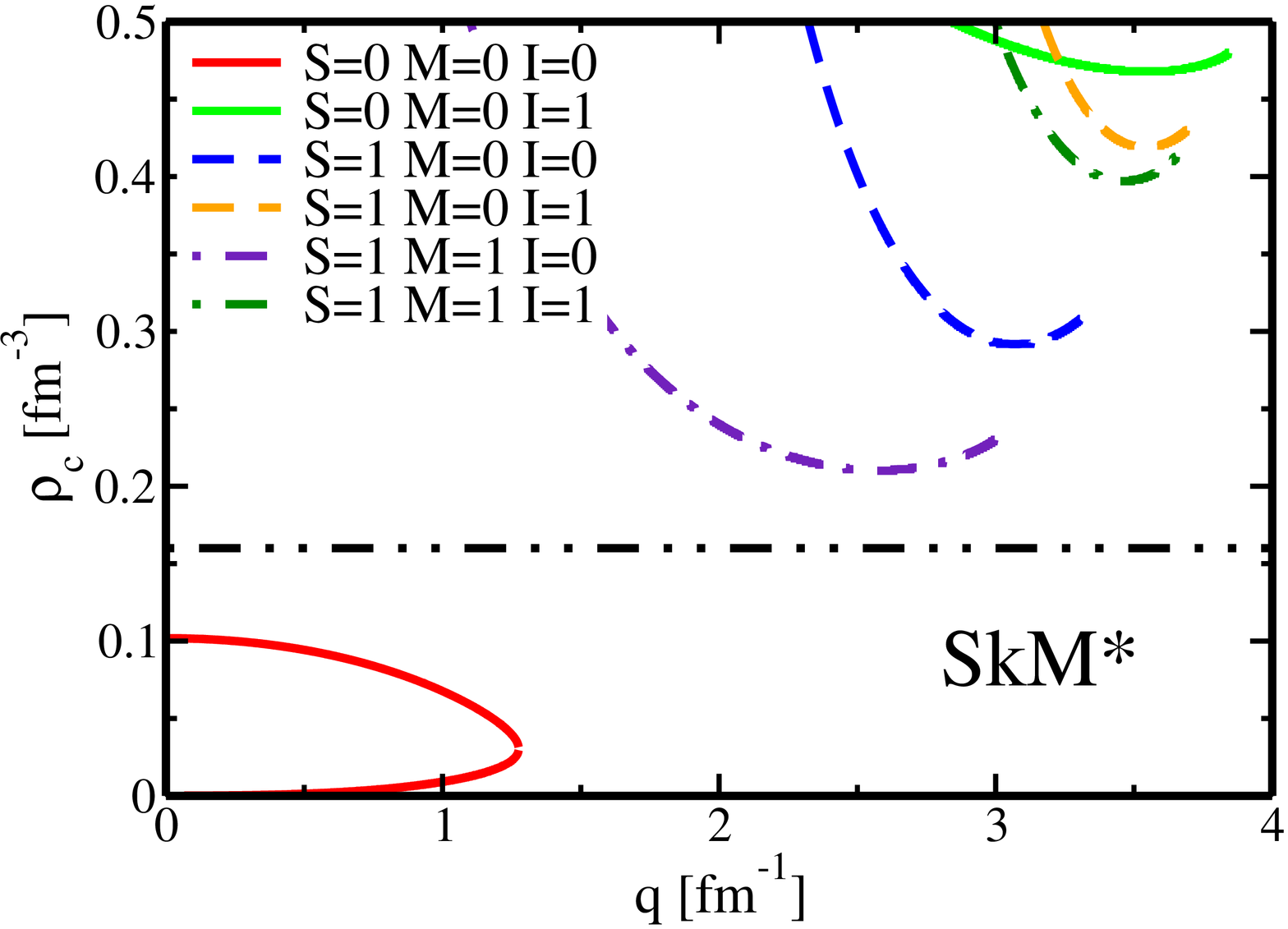}      \\
            \includegraphics[clip,scale=0.3,angle=-90]{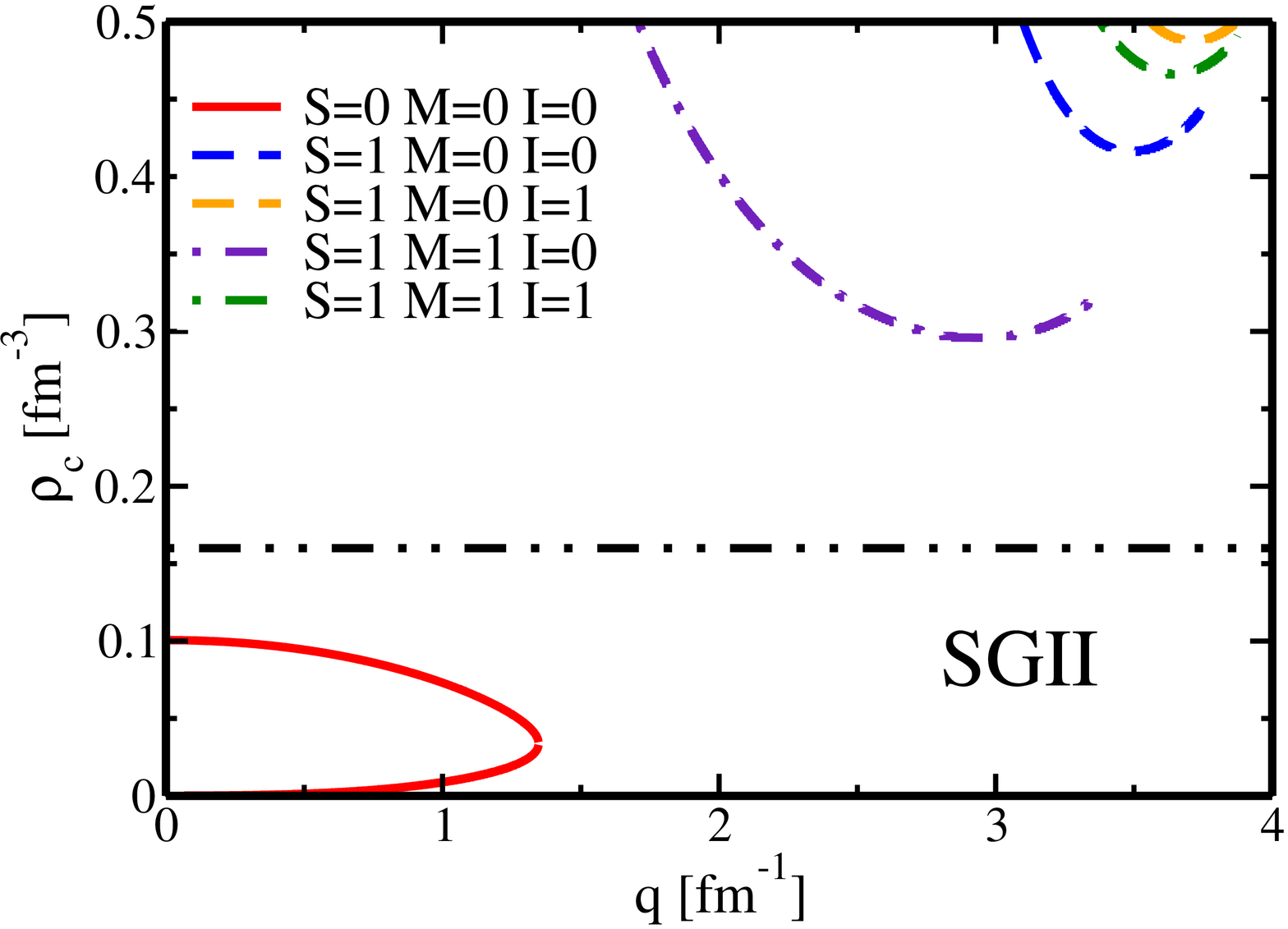}   
      \includegraphics[clip,scale=0.3,angle=-90]{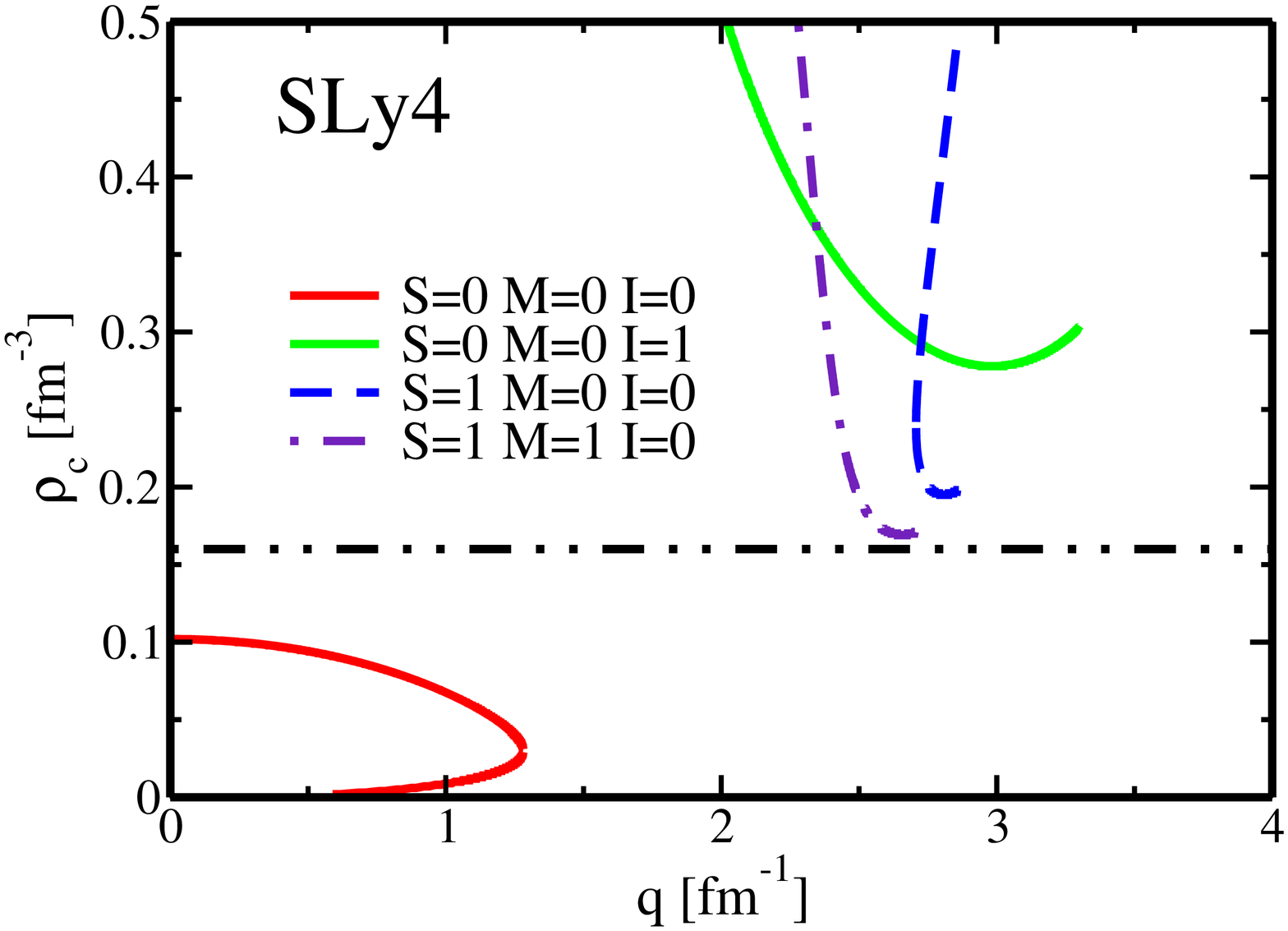}      \\
            \includegraphics[clip,scale=0.3,angle=-90]{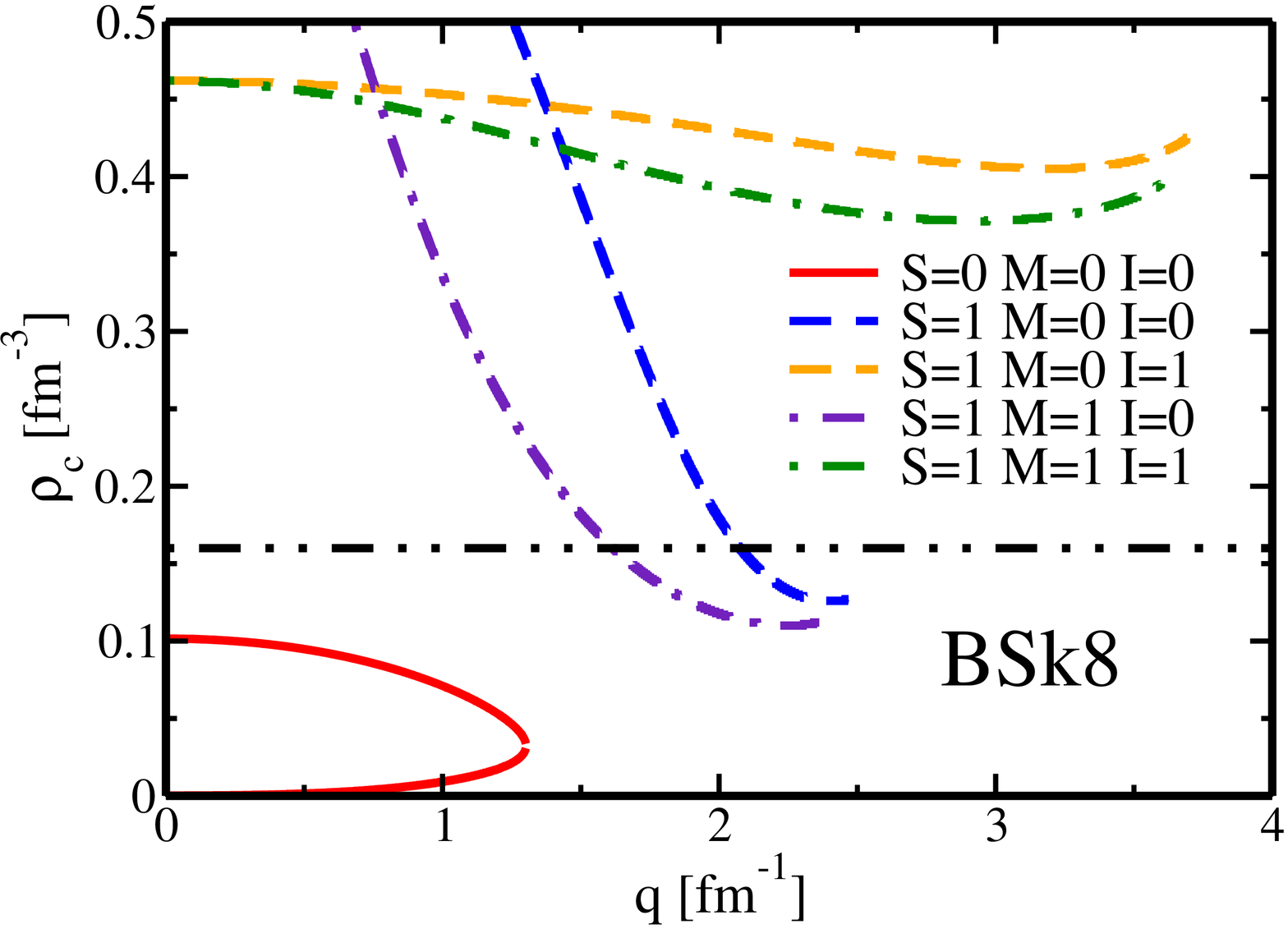}   
      \includegraphics[clip,scale=0.3,angle=-90]{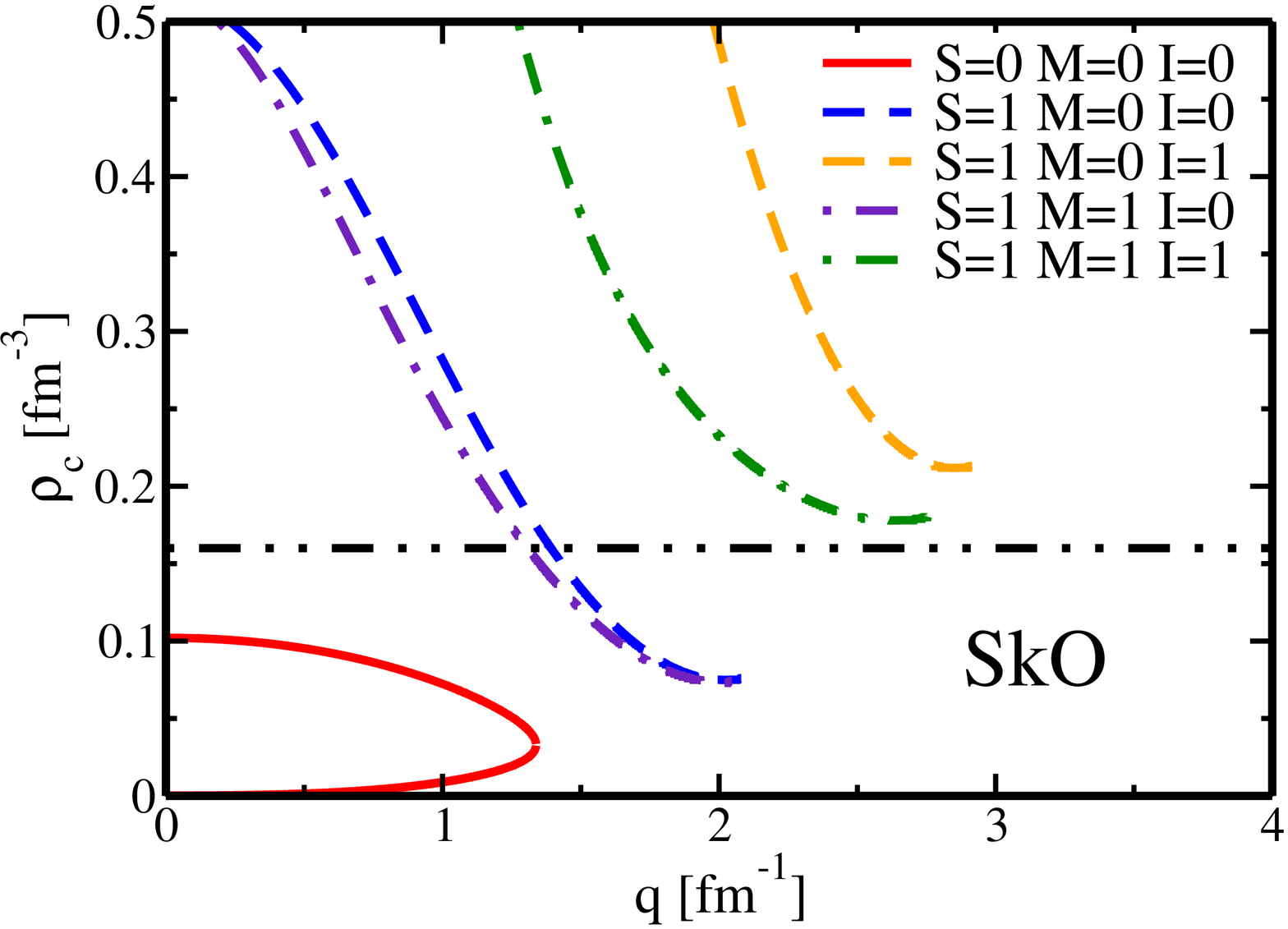}      \\
\ec
      \caption{(Color on line) Critical densities, $\rho_{c}$(in fm$^{-3}$), as 
               functions of the transferred momentum $q$ (in fm$^{-1}$) 
               for some usual Skyrme EDF: 
                SkP~\cite{Dob84a}, SkM*~\cite{Bar82a},   
               SGII~\cite{Giai81}, SLy4~\cite{Cha97a,Cha98a,Cha98b}, 
               BSk8~\cite{Samyn05}, 
               and SkO~\cite{Reinhard99}. The horizontal dashed-dotted line represents the saturation density of the system.}
\label{fig:critical_Skyrme}
\end{figure*}

\begin{figure*}[h]
\bc
      \includegraphics[clip,scale=0.28,angle=-90]{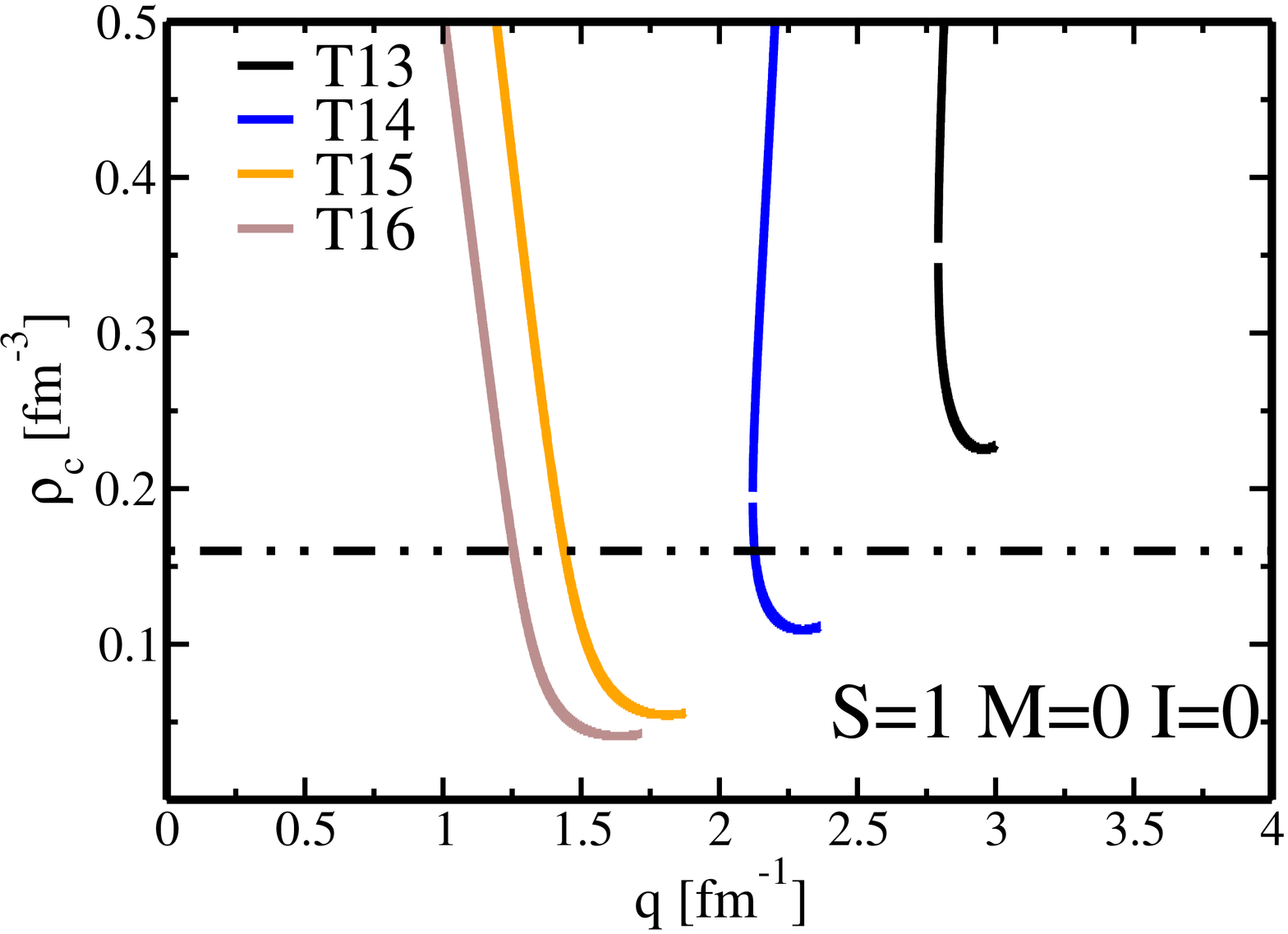}    
            \includegraphics[clip,scale=0.28,angle=-90]{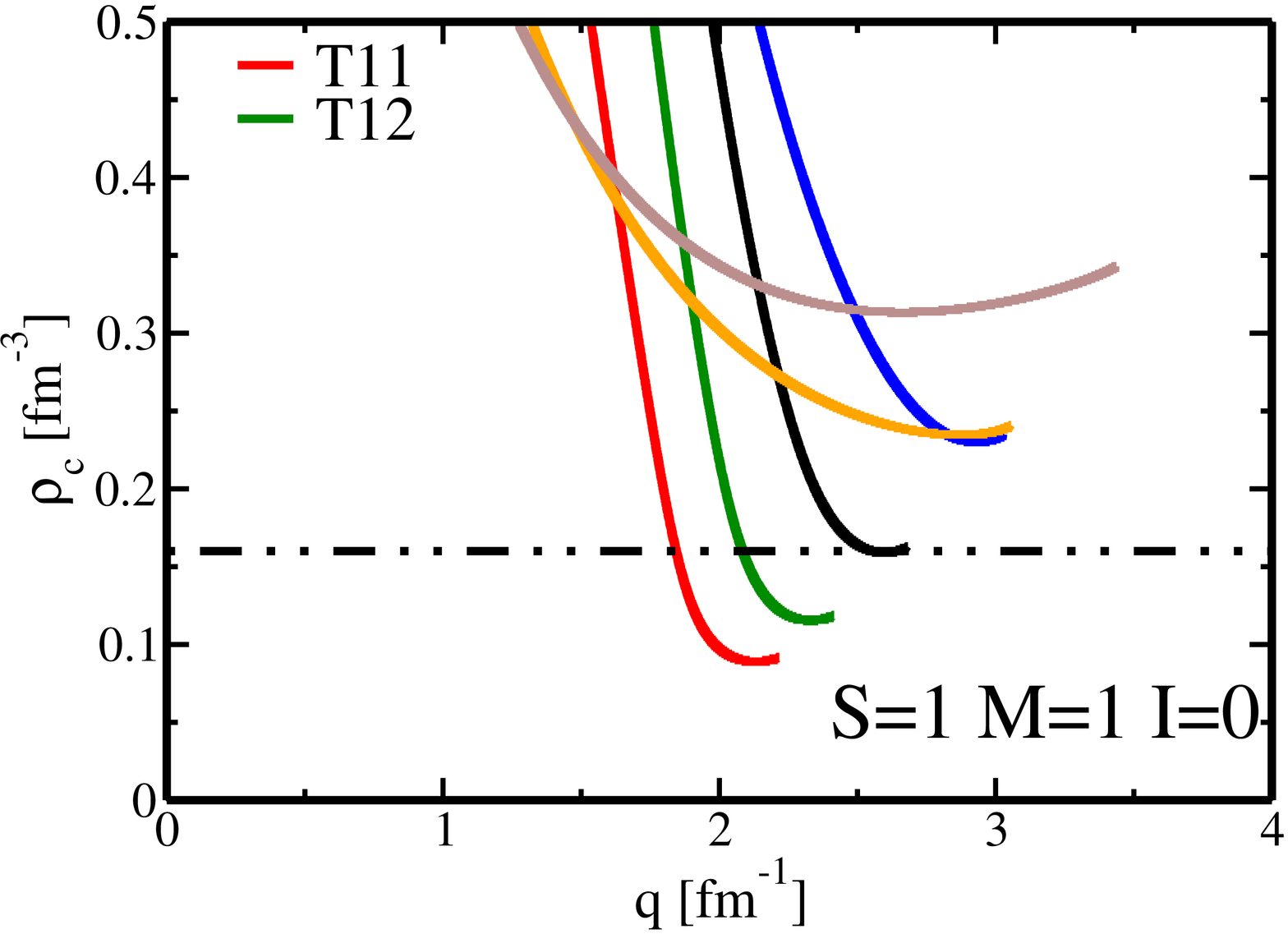}\\    
      \includegraphics[clip,scale=0.28,angle=-90]{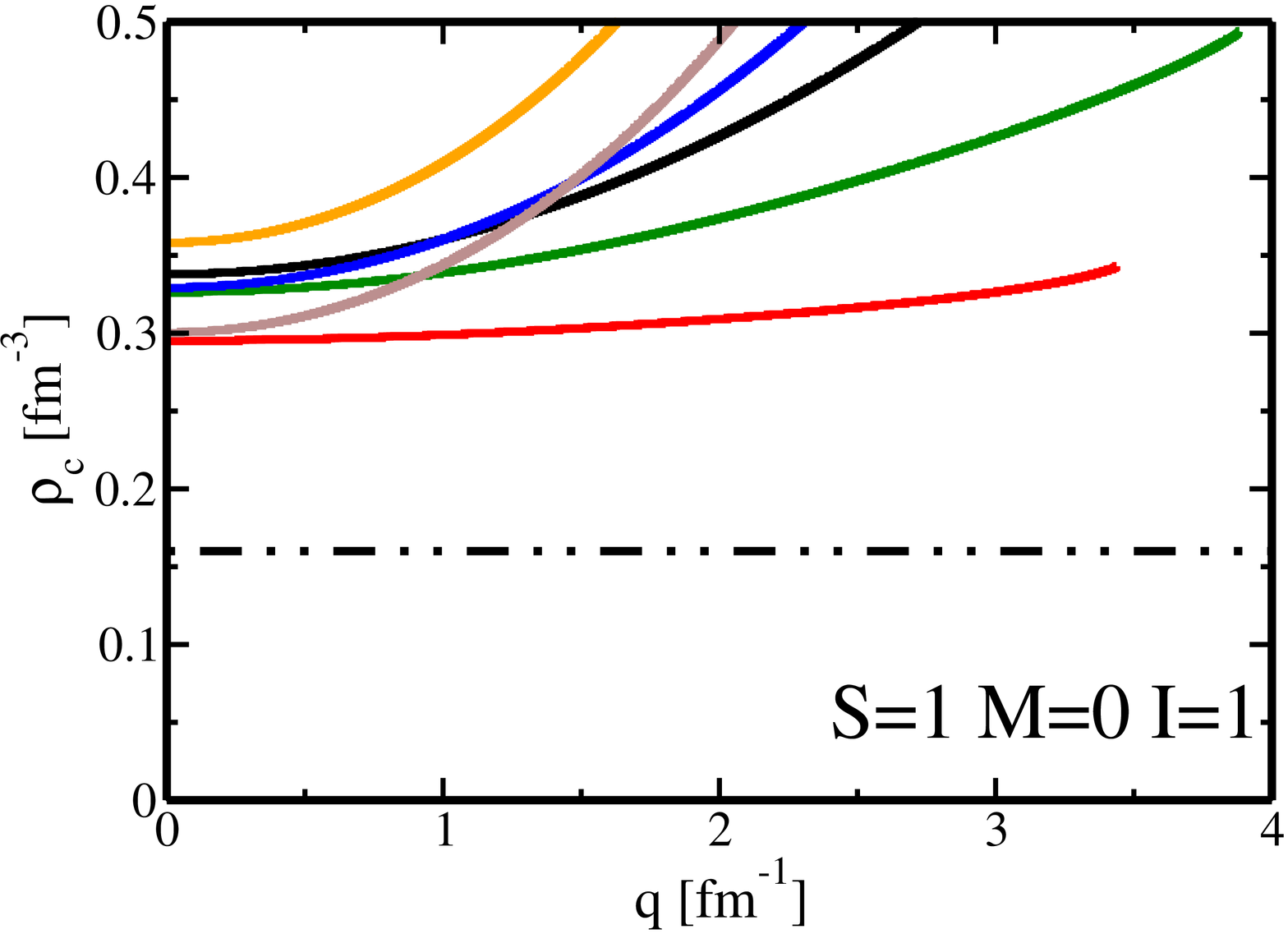}    
      \includegraphics[clip,scale=0.28,angle=-90]{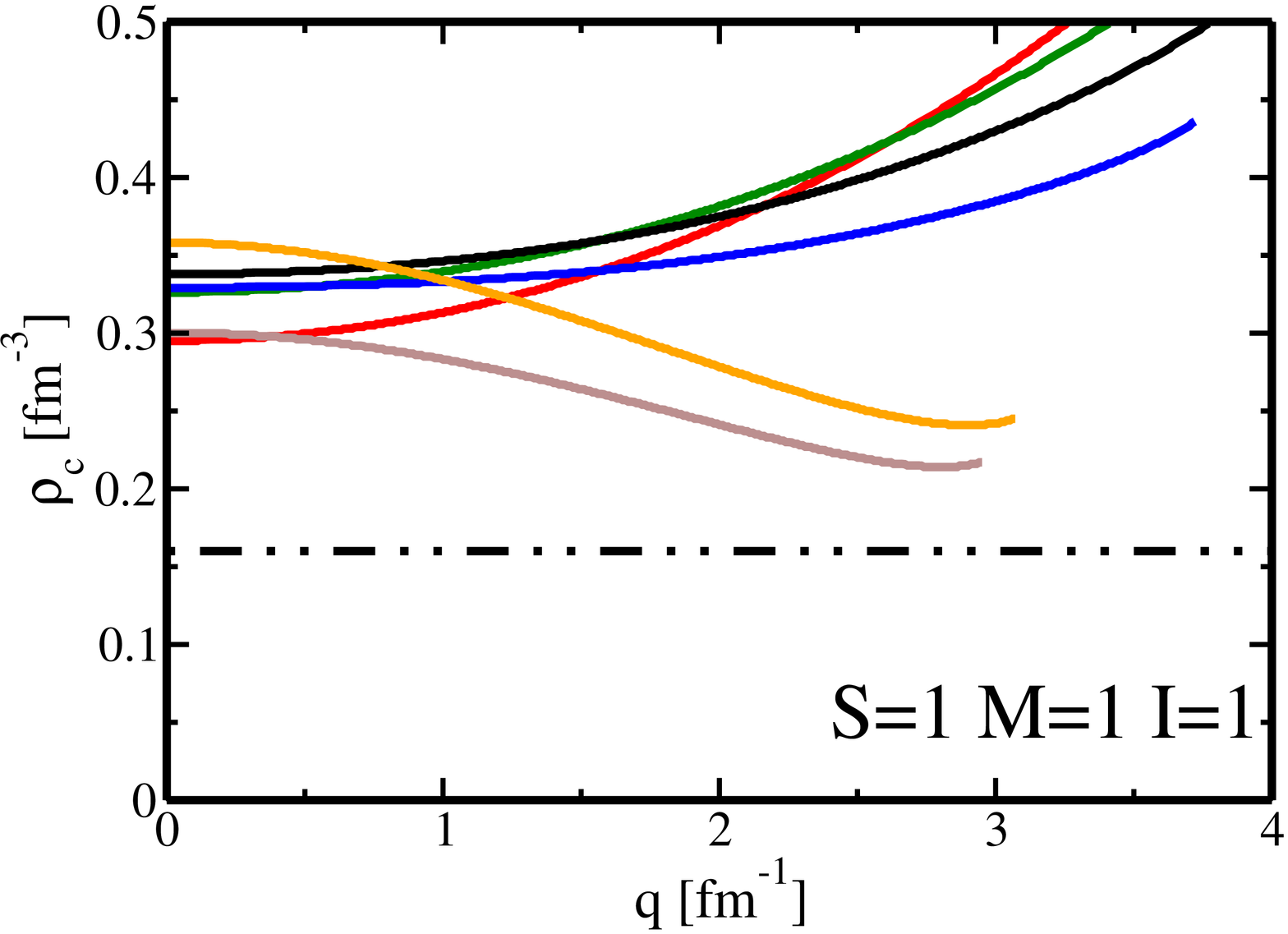}    
      
      \includegraphics[clip,scale=0.28,angle=-90]{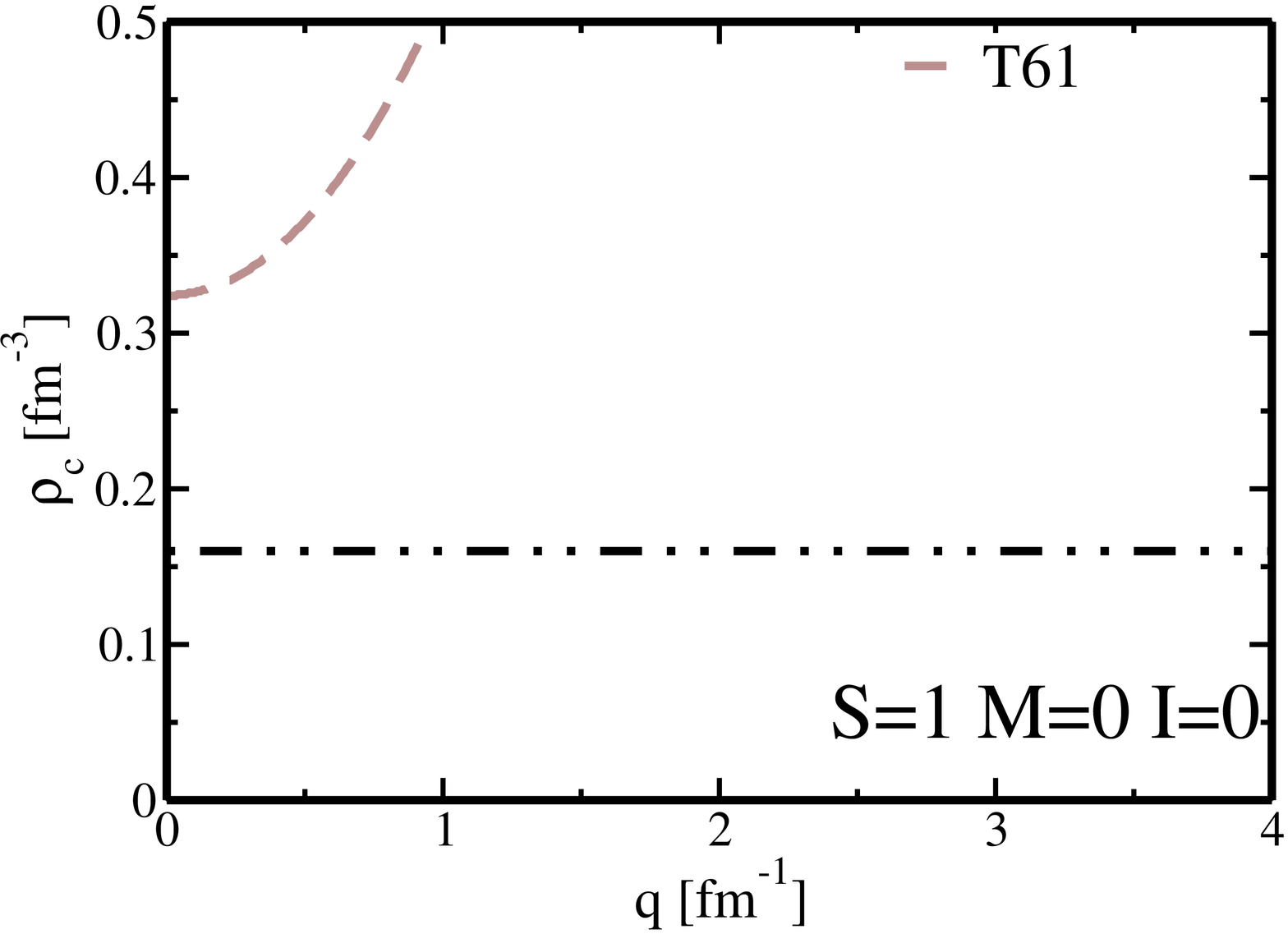}    
            \includegraphics[clip,scale=0.28,angle=-90]{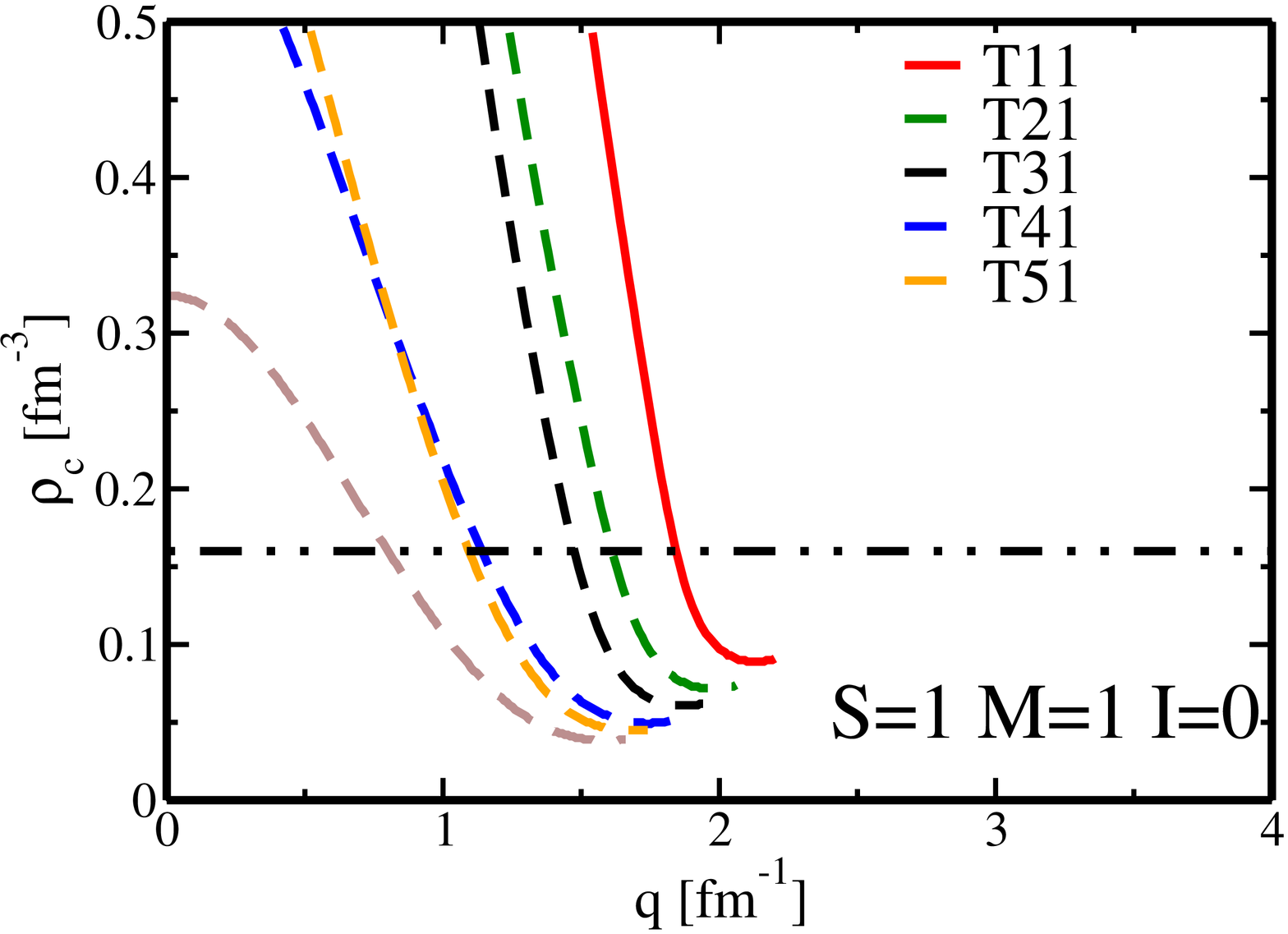}\\    
      \includegraphics[clip,scale=0.28,angle=-90]{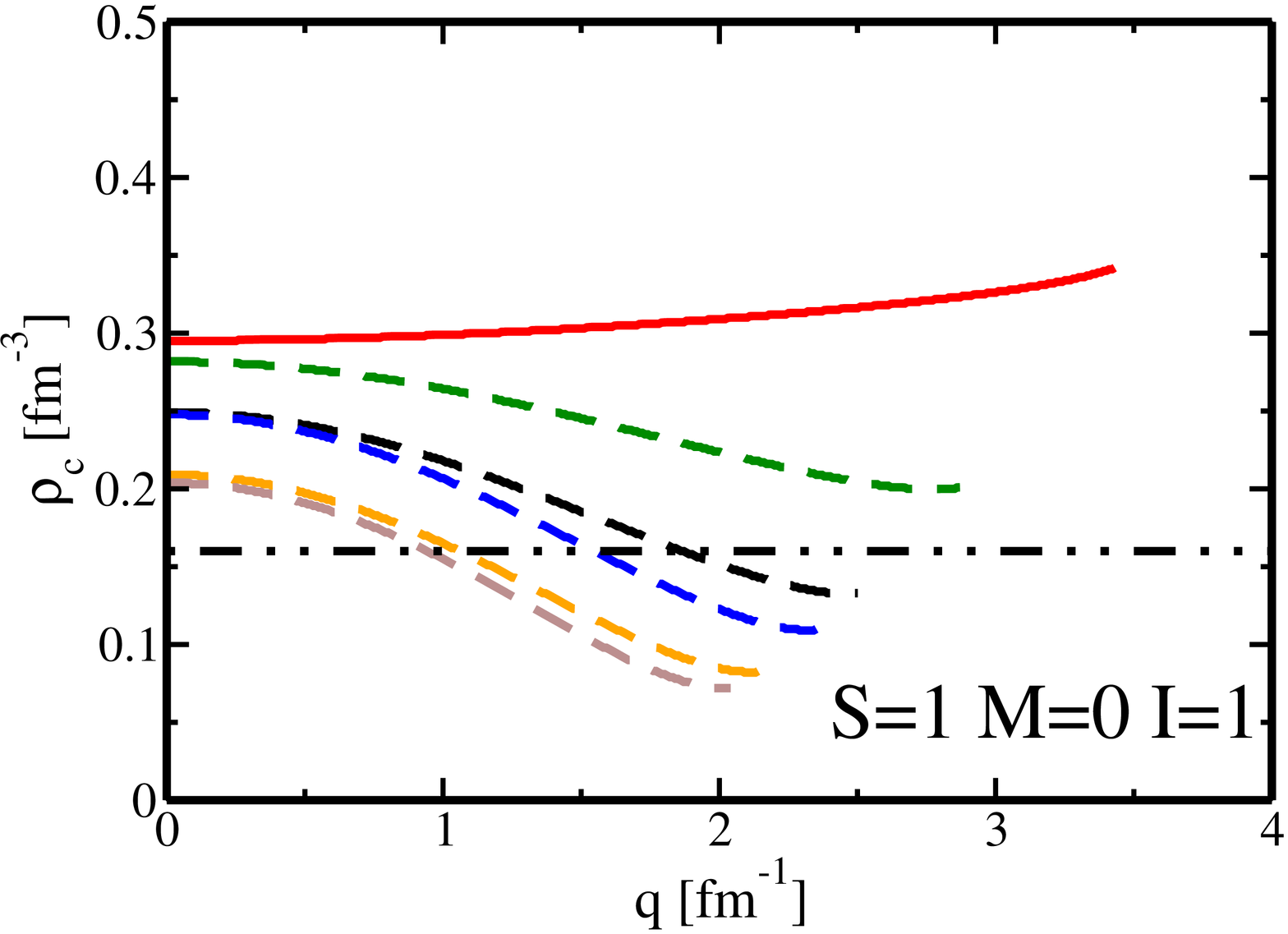}    
      \includegraphics[clip,scale=0.28,angle=-90]{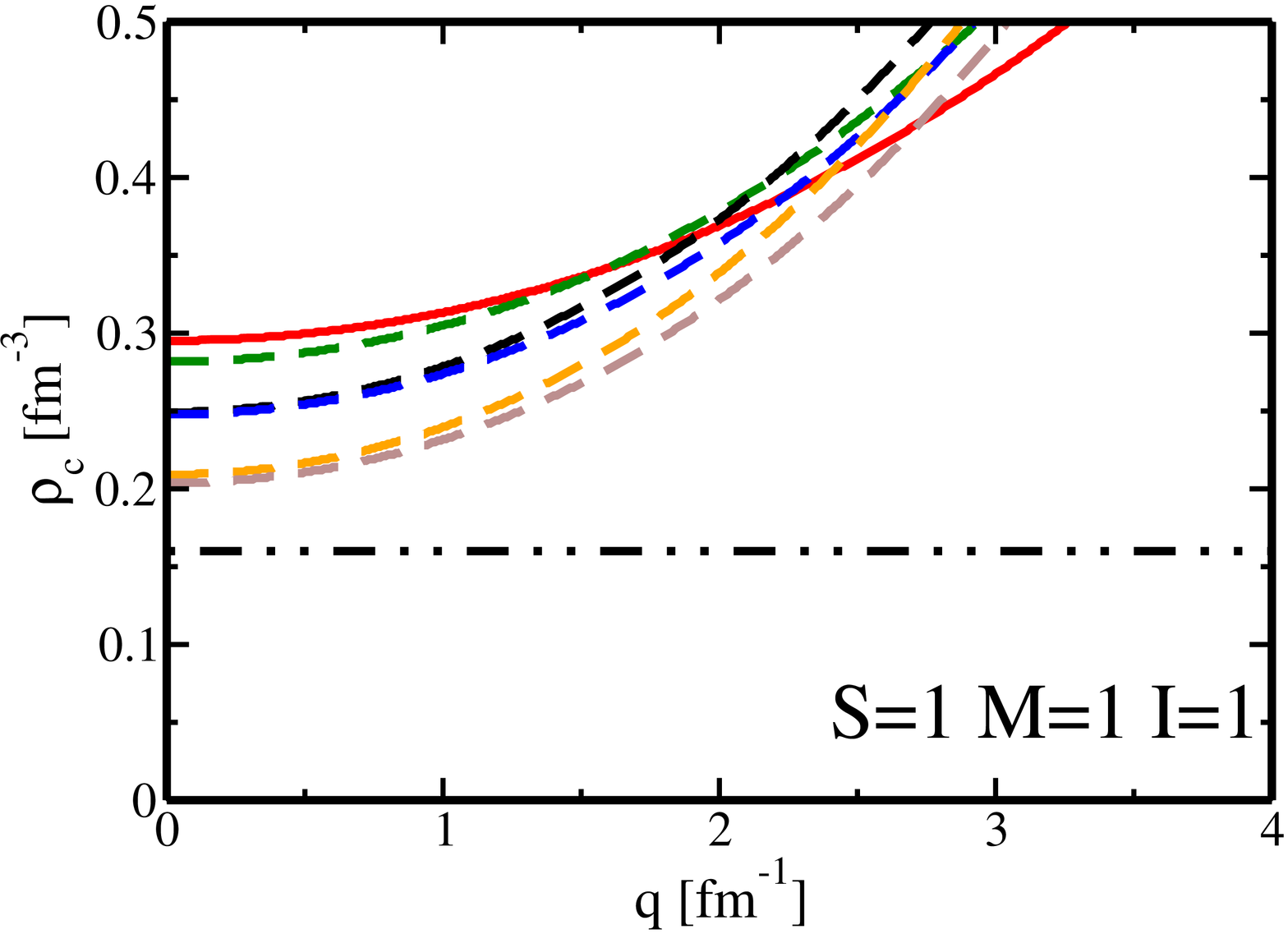}    \\

      \ec
\caption{(Color on line) Critical densities, $\rho_{c}$(in fm$^{-3}$),  in the 
         $S=1$ channels as functions of the transferred momentum 
         $q$ (in fm$^{-1}$) for the T11 to T16 (top four panels) and
         for the T11 to T61 (bottom four panel) tensor parameterizations.The horizontal dashed-dotted line represents the saturation density of the system.}
\label{fig:critical_Seq}
\end{figure*}

Similarly \citeFigure{fig:critical_Seq} shows the evolution of critical 
densities following two series of parameterizations with tensor couplings $T11-T16$ 
and $T11-T61$ when one considers the $C^J_0-C^J_1$ plane of tensor 
coupling constants studied by Lesinski \etal~\cite{Lesinski07}.
In this case we only show the $\text{S}=1$ channel for the different $TIJ$ forces, showing that they all are unstable against spontaneous polarization of finite-size domains as already observed by Hellelmans \emph{et al.} \cite{veerle11}.

\subsection{Landau parameters}

Another important constraint concerning stability of a parametrisation is given by the Landau parameters~: since they represent the short-range of the interaction, they have to be positive. 
The $F_{l},F'_{l}$ spin-independent Landau parameters must obey to the stability condition

\begin{equation}\label{condLandauF}
1+\frac{F_{l}}{2l+1}> 0.
\end{equation}

\noindent Remember that the $l=0$ Landau parametrization  can be related to the second derivative of the EDF with respect to $\mathcal{I},\mathcal{I}_{\tau},\mathcal{I}_{\sigma},\mathcal{I}_{\sigma\tau}$~( see \cite{Jeremy11} for details), the pertinent variables of each $(S,I)$ channel. Eq. (\ref{condLandauF}) represents  thus the condition that the concavity of the equation of state (EoS) at the equilibrium must be positive in each $(S,I)$ channel. The result is represented on \citeFigure{fig:Landau_0_ref} for some Skyrme interactions.
Similarly the $l=1$ Landau parameters can be related to the effective mass in each $(S,I)$ channel and Eq. (\ref{condLandauF}) requires that each effective mass is positive without any pole.
Similar conditions exist for $G_{l},G'_{l}$ spin-dependent Landau parameters. They are shown on  \citeFigure{fig:Landau_1_ref}.
In the presence of a tensor interaction a new additional condition that prevent the deformation of the Fermi sphere has to be satisfied. Following the derivation of Brown \emph{et al.}~\cite{Brown77} we have

\begin{equation}\label{condHlandau}
1+\frac{1}{3}G_{1}-\frac{10}{3}H_{0}> 0 
\end{equation}

\begin{equation}\label{condHlandauB}
1+\frac{1}{3}G_{1}+\frac{5}{3}H_{0}> 0
\end{equation}

\begin{equation}\label{condHlandauC}
1+\frac{1}{3}G_{1}-\frac{1}{3}H_{0}> 0
\end{equation}

\begin{equation}\label{condHlandauD}
\left(1+\frac{G_{0}}{2}\right) + \frac{1}{2} \sqrt{G_{0}^{2}+8H_{0}^{2}} > 0
\end{equation}

\begin{equation}\label{condHlandauE}
\left(1+\frac{G_{0}}{2}\right) - \frac{1}{2} \sqrt{G_{0}^{2}+8H_{0}^{2}} > 0
\end{equation}

\noindent and similarly for  $H'_{0}$. On  \citeFigure{fig:Landau_H_ref} we show the left hand side of Eqs. (\ref{condHlandau}-\ref{condHlandauE}).
 This result is consistent with the results presented by Cao \emph{et al.} \cite{Colo10}, but generalized here for the case of a Skyrme functional.
\bc
\begin{figure*}[htbp]
\bc
      \includegraphics[clip,scale=0.3,angle=-90]{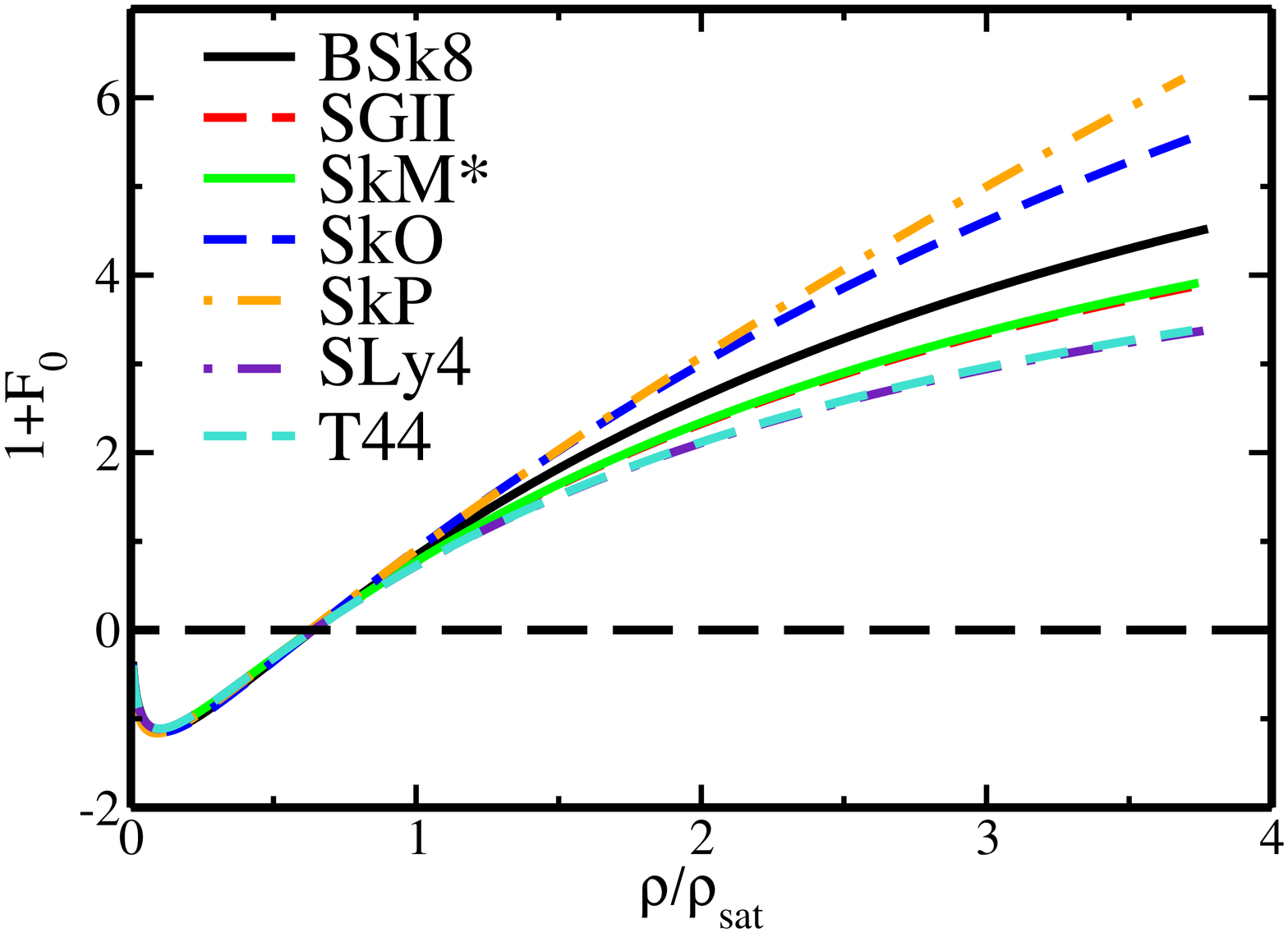}    
      \includegraphics[clip,scale=0.3,angle=-90]{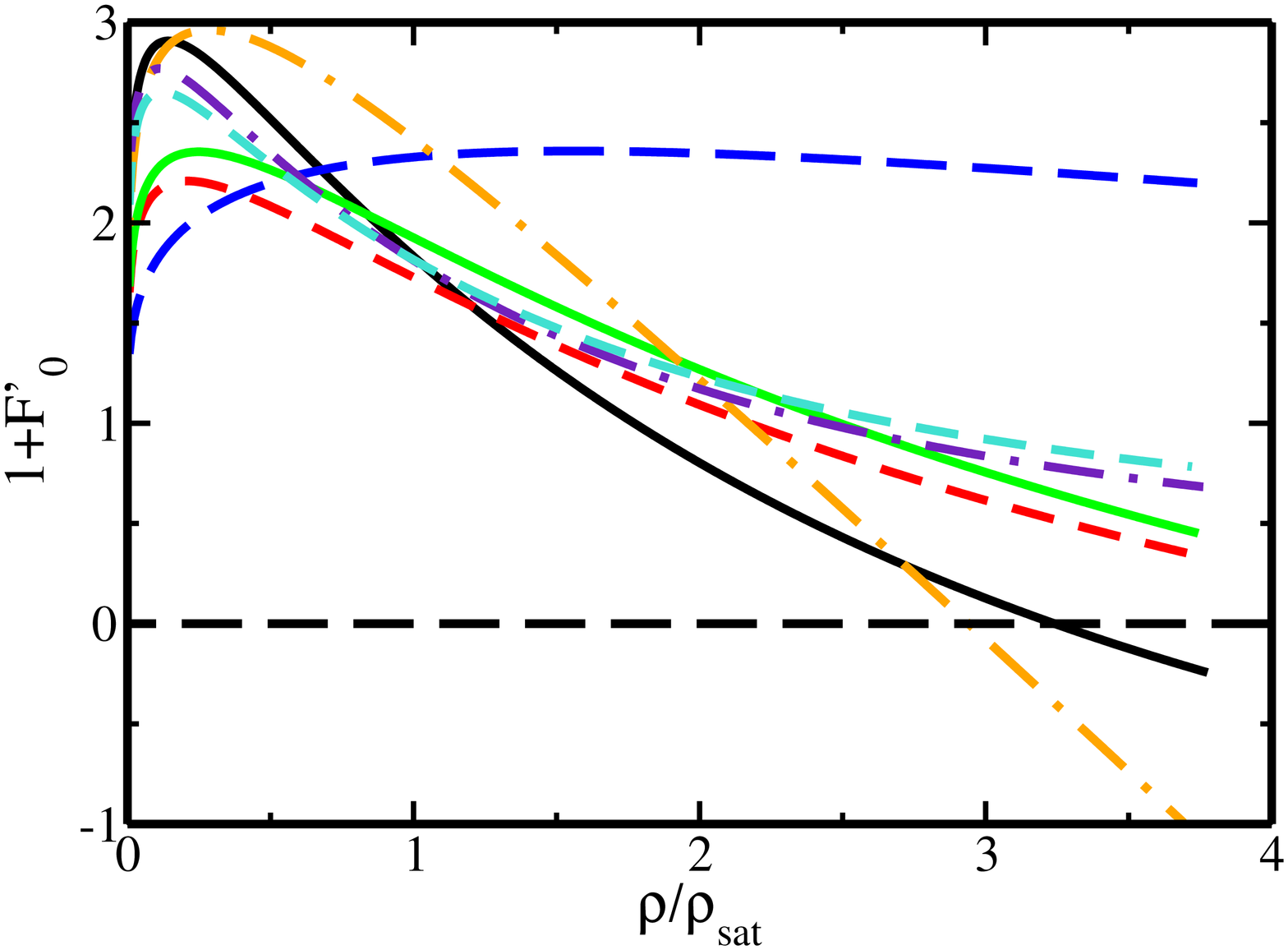}\\    
      \includegraphics[clip,scale=0.3,angle=-90]{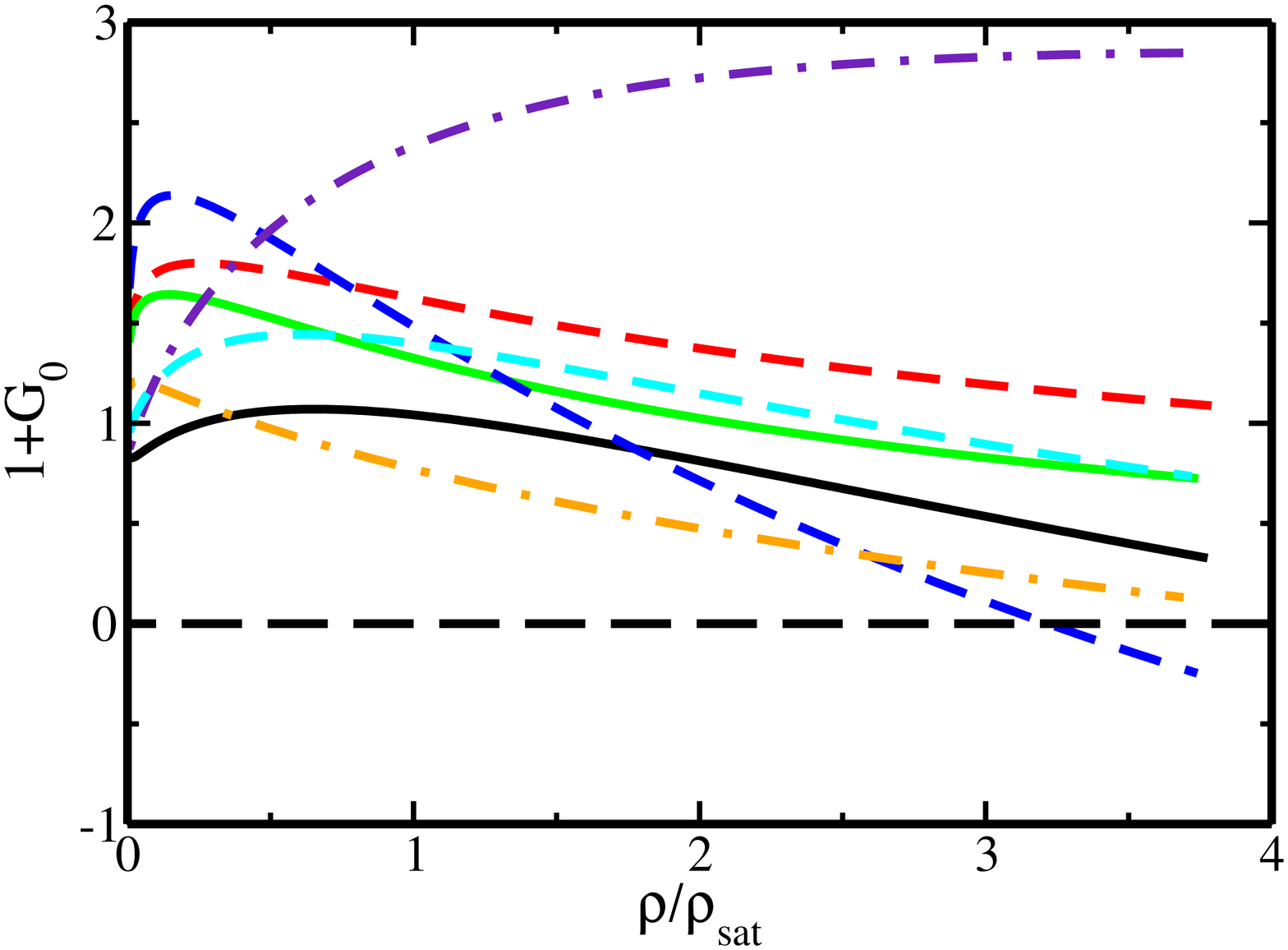}    
      \includegraphics[clip,scale=0.3,angle=-90]{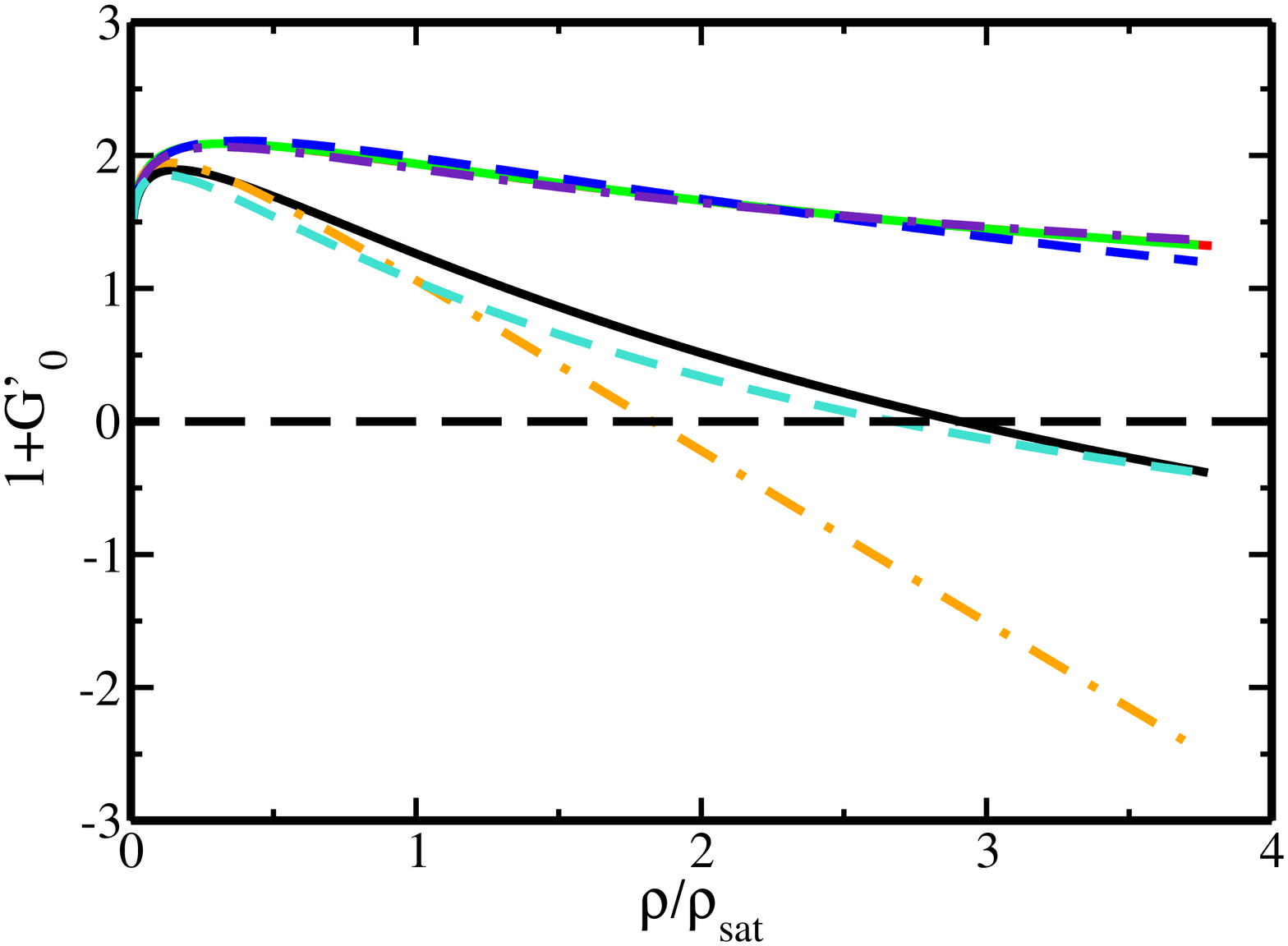}   
\ec 
                \caption{$\ell=0$ Landau parameters as a function of the ratio $\rho/\rho_{sat}$, where $\rho$ is the density of the system and $\rho_{sat}$ is the saturation density,
                for some usual Skyrme parameterizations.}
       \label{fig:Landau_0_ref}
\end{figure*}

\begin{figure*}[htbp]
     \bc
      \includegraphics[clip,scale=0.3,angle=-90]{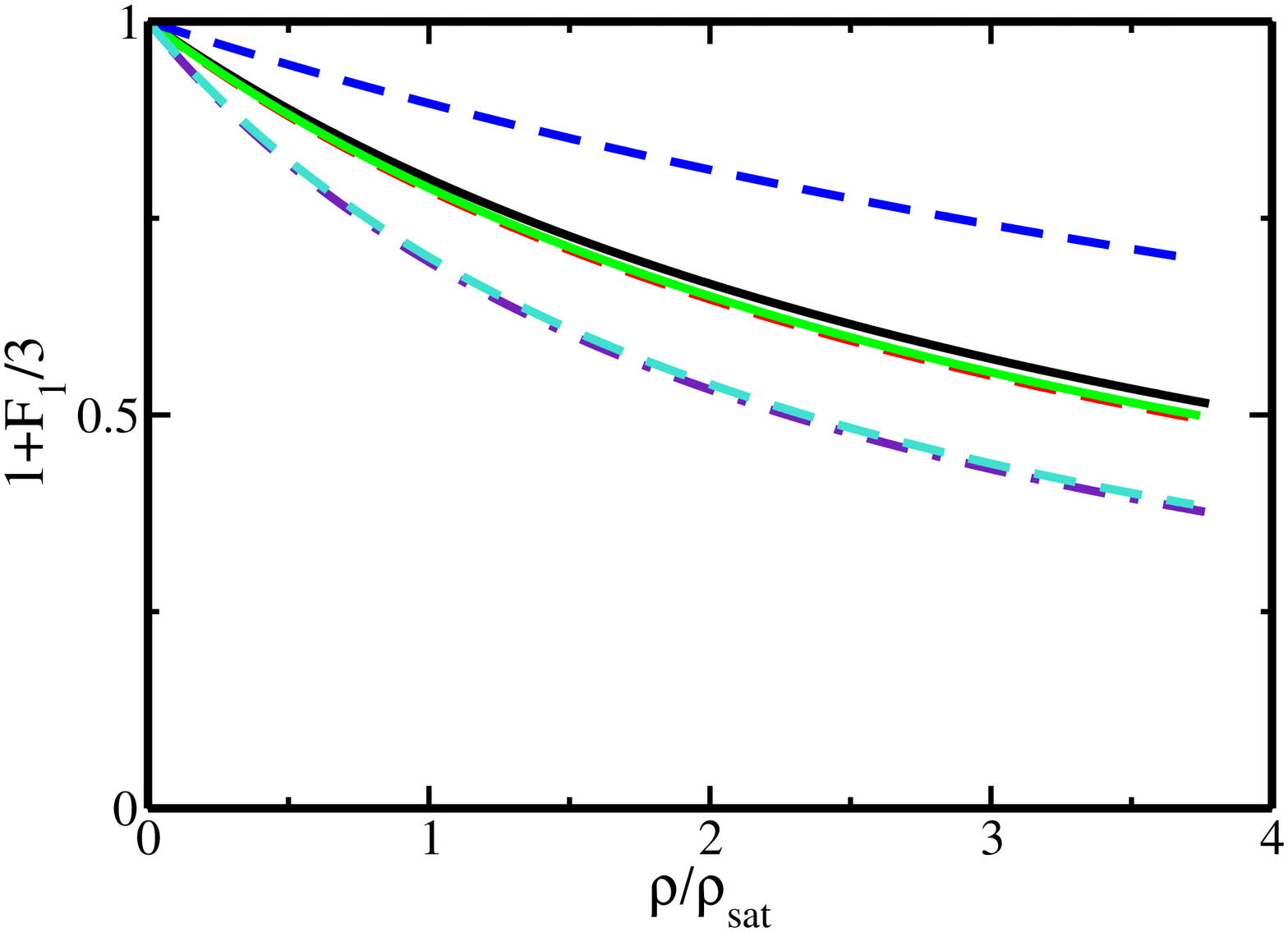}    
      \includegraphics[clip,scale=0.3,angle=-90]{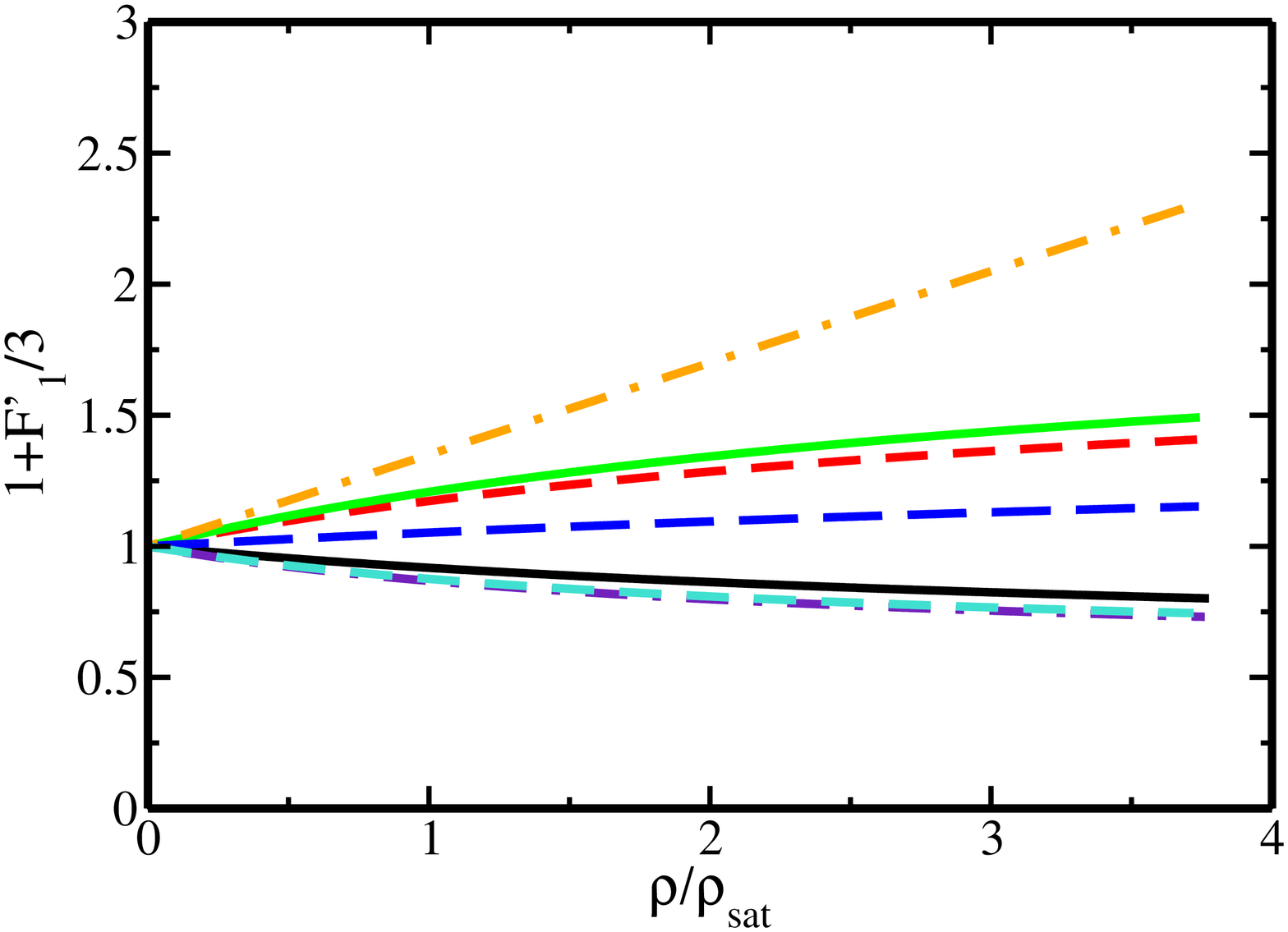}\\    
      \includegraphics[clip,scale=0.3,angle=-90]{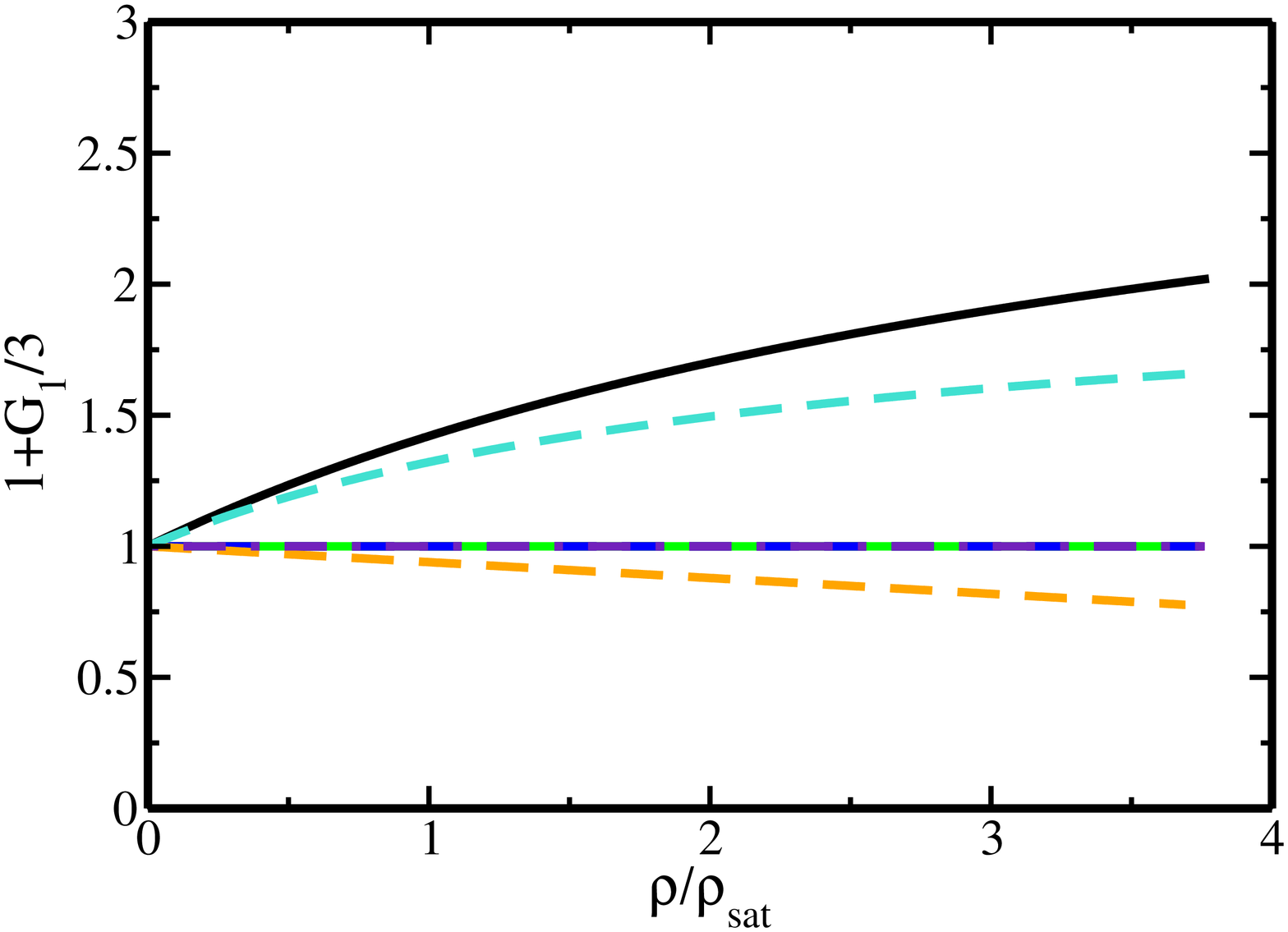}    
      \includegraphics[clip,scale=0.3,angle=-90]{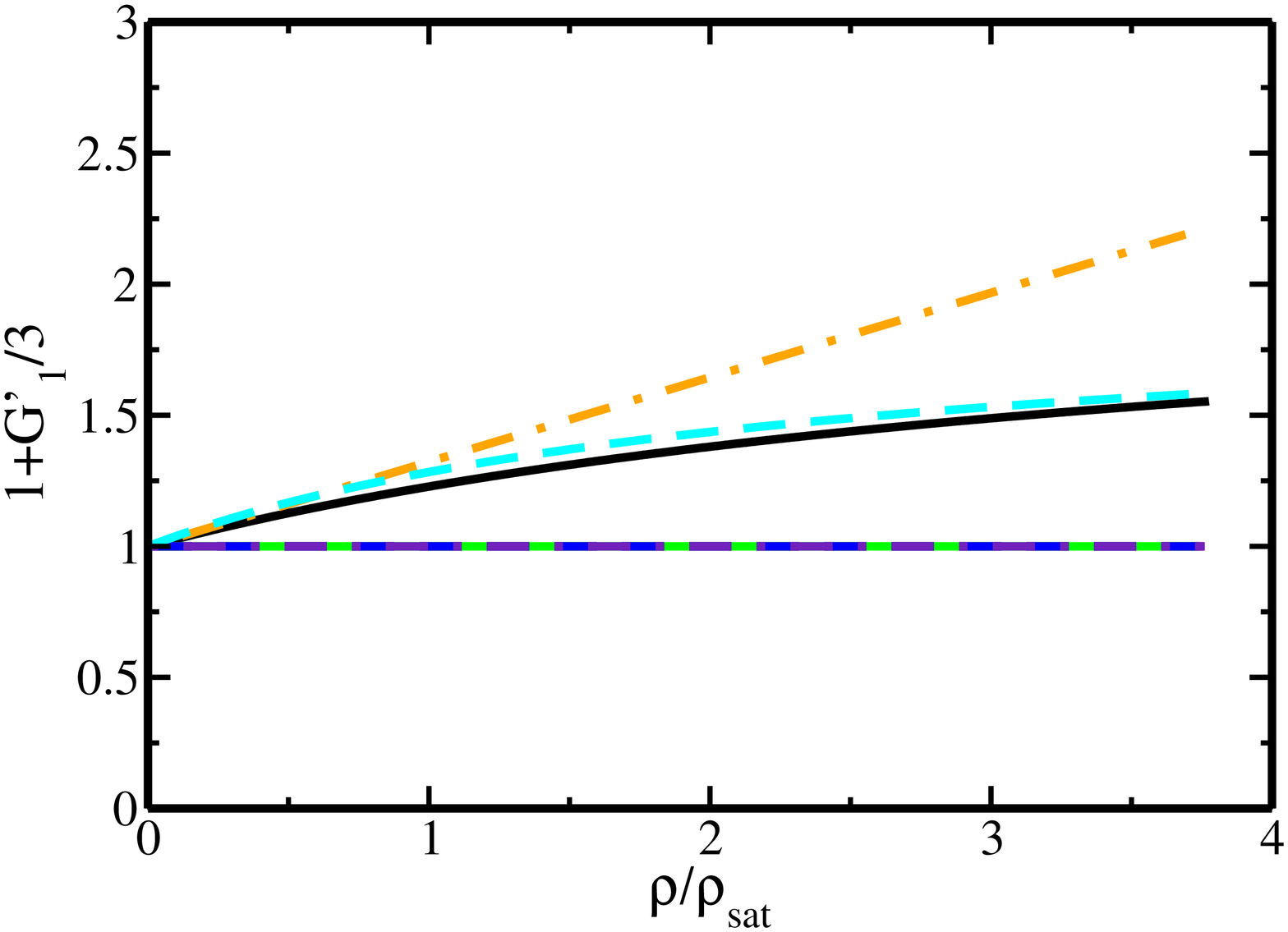}   
\ec 
      \caption{$\ell=1$ Landau parameters as a function of the ratio $\rho/\rho_{sat}$, where $\rho$ is the density of the system and $\rho_{sat}$ is the saturation density, for some usual Skyrme parameterizations. See \ref{fig:Landau_0_ref} for other details.}
       \label{fig:Landau_1_ref}
\end{figure*}

\begin{figure*}[htbp]
     \bc
      \includegraphics[clip,scale=0.3,angle=-90]{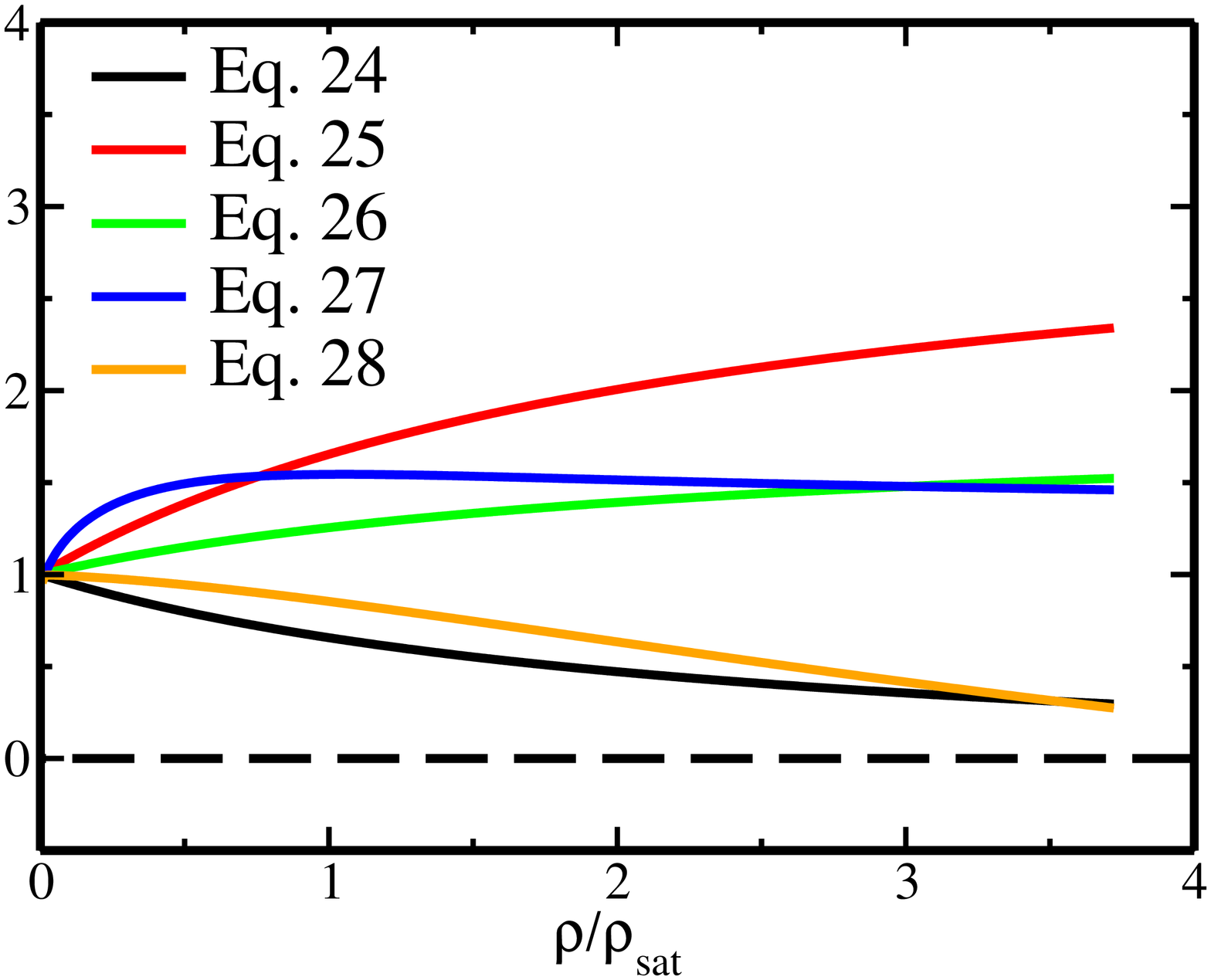}    
      \includegraphics[clip,scale=0.3,angle=-90]{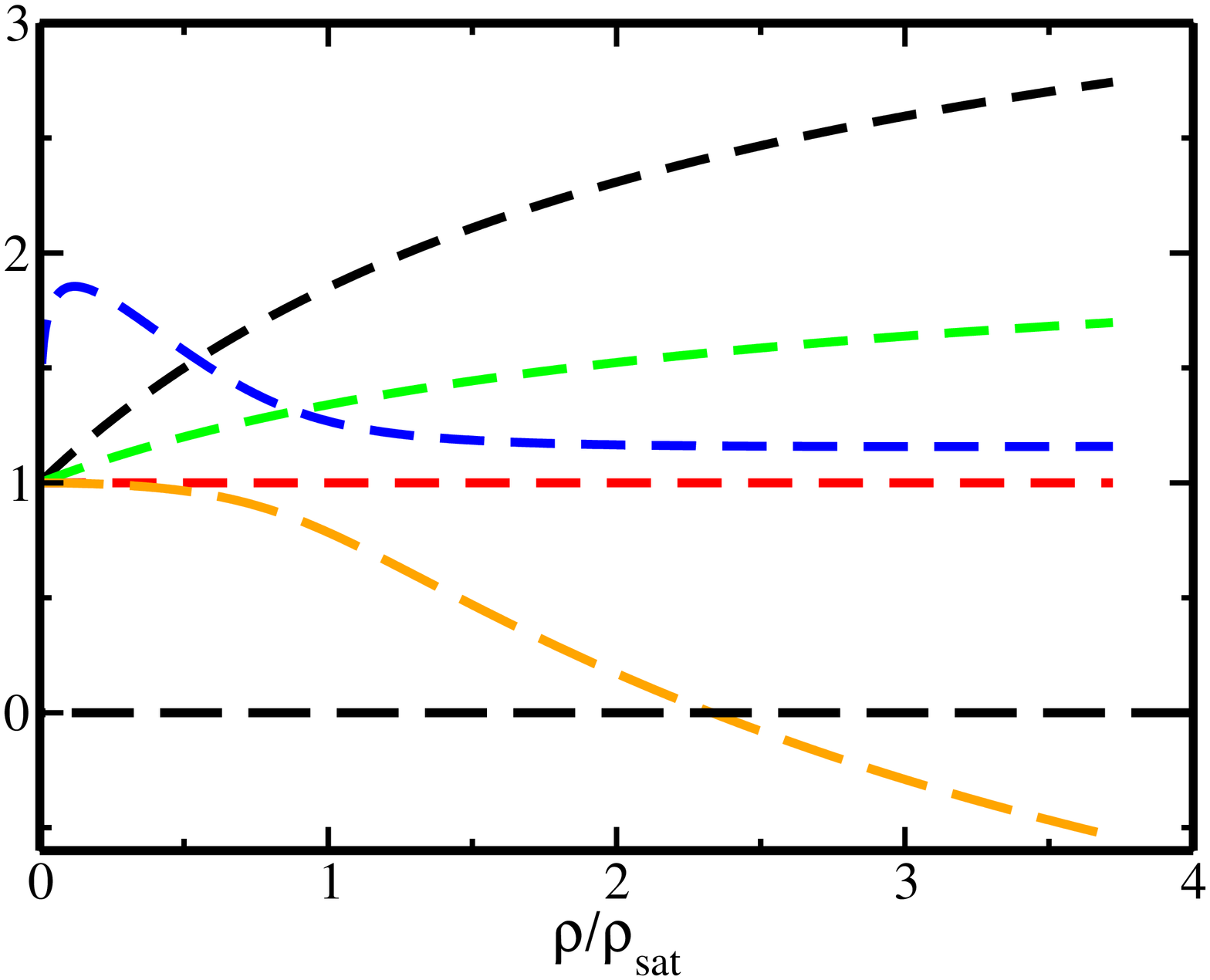}
\ec 
      \caption{On the left panel we show the left hand side of Eqs. (\ref{condHlandau}-\ref{condHlandauE}), while on  the right panel, adopting the same color code, the same but replacing in Eqs. (\ref{condHlandau}-\ref{condHlandauE}) with the terms $G_{0}',G_{1}',H_{0}'$. See text for details.}
       \label{fig:Landau_H_ref}
\end{figure*}
\ec

%
\section{Summary and conclusions}
\label{sect:conclusions}

In this article, we have presented the analytic contribution arising from the tensor terms to the RPA response functions with a general EDF as a starting point. From these response functions, we derived the Landau parameters and we  focused on instabilities at zero energy and finite transfer momentum.
In particular, we have shown that a divergence of the response structure functions $\chi^{(\alpha)}(\mathbf{q},\omega)$
indicates a finite size instability in infinite matter. Moreover this instability can be detected by simply using the analytical IEWSR, which is a great advantage for future applications.
At this point, one also has to note that a systematic study of the critical
densities is in progress in order to determine whether the link
between the divergences of  $\chi^{(\alpha)}(\mathbf{q},\omega)$ and the instabilities
encountered in nuclei at the Hartree-Fock approximation is robust.

Another important point under study is the identification, directly from
the Skyrme energy functional, of the contribution of each term of the EDF
in the response functions. Such a study would enable us to put some constraints on the different constants in order to avoid instabilities.
In the same spirit, a detailed study of sum rules can enlighten the
contribution of the tensor
for various physical situations (see for instance \cite{Lipparini}). 
Finally, applications to pure neutron matter is of great importance (see for
instance~\cite{Marcos91,Fantoni01,Margueron02,Vidana02a,Vidana02b,
Margueron09,Isayev04,Beraudo04,Beraudo05,Rios05,Lopez-Val06,Krastev07,
Bordbar08}) and will be the subject of a forthcoming article in preparation.
In that case, the above formulae are no longer directly usable and have 
been adapted to that specific case.


\section*{Acknowledgments}

This work was supported by the NESQ project (ANR-BLANC 0407).
The authors thanks C. Ducoin for enligthening discussions
about the spinodal instability and also J. Margueron, J. Navarro, T. Duguet, M. Bender and M. Ericson
for pertinent comments on this work.
M.M. acknowledges the Communaut\'e Fran\c caise de Belgique (Actions de Recherche Concert\'ees) for financial support.


\clearpage

\begin{appendix}

%
\section{Particle-hole matrix elements of the zero range tensor part of the 
         interaction.}
\label{app:phme}

Following the notation adopted in article I, we give in Table~\ref{arrayfinale} the values of the particle-hole residual interaction for the tensor part of the functional.
\bwt

\begin{table}[!h]
\caption{Contribution of the EDF tensor part to the residual interaction
 in terms of the $B_I$ coupling constants. 
For the sake of simplicity we have introduced the notation:
$\bbK_{\rm i,j}=[(k_{12})_{i}(k_{12})_{j}]$, 
where $(k_{12})_{M}^{(1)}$ is defined in Eq.(9) of article I. 
The term $\delta_{SS'}\delta_{S1}\delta_{II'}\delta_{QQ'}$ is implicit everywhere.}
\begin{ruledtabular}
\begin{tabular}{cccc} 
                     &&&                                                               \\
                     & $M'=1$ & $M'=0$ & $M'=-1$                                       \\ 
                     &        &        &                                               \\ 
\hline
                     &        &        &                                               \\
\multirow{3}{12mm}{$M=1$} & $-2 q^2 \, \left( B^T_I + 4 B^{\Delta s}_I \right)$                               
                          &&                                                           \\ 
                          & $+4 \, B^T_I \, \bbKzz$               
                          & $-4 \, B^F_I \, \bbKmz$  
                          & $-4 \, B^F_I \, \bbKmm$                                    \\ 
                          & $-4 \, \left( 2 B^T_I + B^F_I \right) \, \bbKum$              
                          &&                                                           \\
                          &&&                                                          \\ 
\hline
                          &&&                                                          \\ 
\multirow{3}{14mm}{$M=0$} && $-2 \left( B^T_I - 4 B^{\nabla s}_I + 4 B^{\Delta s}_I + B^F_I \right) q^{2}$
                          &                                                            \\     
                          & $ 4 \, B^F_I \, \bbKzu$    
                          & $+4 \, \left( B^T_I + B^F_I \right) \, \bbKzz$   
                          & $ 4 \, B^F_I \, \bbKmz$                                    \\
                          &&$-8 \, B^T_I \, \bbKum$   
                          &                                                            \\
                          &&&                                                          \\ 
\hline
                          &&&                                                          \\   
\multirow{3}{14mm}{$M=-1$}&&& $-2 q^2 \, \left( B^T_I + 4 B^{\Delta s}_I \right)$      \\ 
                          & $-4 \, B^F_I \, \bbKuu$         
                          & $-4 \, B^F_I \, \bbKuz$            
                          & $+4 \, B^T_I \, \bbKzz$                                    \\
                          &&& $-4 \, \left( 2 B^T_I + B^F_I \right) \, \bbKum$         \\
                          &&&                                                          \\

\end{tabular}
\label{arrayfinale}
\end{ruledtabular}
\end{table}

\ewt

%
\section{RPA nuclear responses}
\label{app:responses}

We recall here the nuclear responses already given in article I
but rewritten here in terms of the coupling constants of the Skyrme EDF.
We keep in mind from article I the definitions of these coupling 
constants in terms of the parameters of the Skyrme interaction.

\bwt

\bi
   \item For the $S=0$ channel

\bqr
\label{S0snm_tensor}
\frac{\chihf}{\chi^{{\rm (0,I)}}_{RPA}} 
 & = & \, 1 \, - \, \wwwuzI \chiz 
                       \, + \, \wdzI \left( \pdemi q^2 \chiz - 2 k_F^2 \chid \right)                           \nn \\    
 & + & \left[ \wdzI \right]^2 \left[ k_F^4 \chidd - k_F^4 \cchi 
                                   + m^{*2} \omega^2 \chizd - \tfrac{1}{4} q^2 \mrho \chiz \right]  \nn \\
 & + & \, 2 \moqd \frac{\wdzI}{1-\pdemi \mrho \wdzI } \; \chiz        \q , \\
\nn  
\eqr

\noi
with

\be
\wwwuzI \, = \, \wuzI
            + 16 \; q^4 \, \left[ C_I^{\nabla J} \right]^2 \, 
            \frac{\bdbt}{1 + q^2 \pbdbt \left[ \wdzI  + 4 B_I^T - 2 B_I^F \right]}   \q . 
\ee

   \item and for the $S=1$ channels

\bqr
\label{responseM0S1}
\frac{\chi_{HF}}{\chi_{RPA}^{{\rm (1,0,I)}}} & = &
  \left[ 1 + \mrho B_I^F \right]^2 \, + \, \wwwuuzI \chiz   \nn \\
                                             & + & 
  \left[ \wduI + 4 B_I^T \right ] \left\{ \pdemi q^2 \left[ 1 + 2 \mrho B_I^F \right] \chiz 
                - 2k_F^2 \chid + 2 \mrho k_F^2 B_I^F \left( \chiz - \chid \right)                                                                     \right\}                          \nn \\
                                             & + & 
  \left[ \wduI + 4 B_I^T \right]^2 \left\{ 
  k_F^4 \chid^2 - k_F^4 \chiz \chiq + m^{*2} \omega^2 \chiz^2 - \tfrac{1}{4} q^2 \mrho \chiz      
                                   \right\}                         \nn \\
                                             & + & 
  2 \chiz \, \moqd \,
  \frac{\left[ \wduI  + 4 B_I^T + 4 B^F_I \right] \left[ 1 + \pdemi \mrho X^{{\rm (1,0,I)}} \right]} 
       {1 - \pdemi \mrho \left[ \wduI  + 4 B_I^T + 4 B^F_I - X^{{\rm (1,0,I)}}\right]}
                                                                \q ,  \\
\nn                                                                                   
\eqr

\bqr
\label{responseM1S1}
\frac{\chi_{HF}}{\chi_{RPA}^{{\rm (1,\pm1,I)}}} & = & 
  \left[ 1 - \pdemi \mrho B_I^F \right]^2 - \wwwuuuI \chiz            \nn \\
                                                & + &                                                
  \left[ \wduI + 4 B_I^T + 2 B_I^F \right] \left\{ \pdemi q^2 \left[ 1 - \mrho B_I^F \right] \chiz 
                                        - 2 k_F^2 \chid - \mrho k_F^2 B_I^F \left( \chiz - \chid \right)  
                                           \right\}                                            \nn \\
                                                & + & 
  \left[ \wduI + 4 B_I^T + 2 B_I^F \right]^2 \left\{
   k_F^4 \chid^2 -  k_F^4 \chiz \chiq + m^{*2} \omega^2 \chiz^2 - \tfrac{1}{4} \mrho q^2 \chiz   
                                             \right\}                                         \nn \\
                                                & + &
  2 \chiz \, \moqd \, 
  \frac{\left[ \wduI + 4 B_I^T \right] \left[ 1 + \frac{1}{4} \mrho X^{{\rm (1,\pm1,I)}} \right]} 
       {1 - \demi \mrho \left[ \wduI  + 4 B_I^T - \demi X^{{\rm (1,\pm1,I)}} \right]}   \q ,  \\
\nn  
\eqr

\noi
were we have used

\bqr
\wwwuuzI & = & - \left[ \wuuI + 8 q^2 \left( B_I^{\nabla s} - B_I^{\Delta s} \right) \right]
               + 2 q^2 B_I^T                                                             \nn \\
         & + & \left[ 2 q^2 - 8 \moqd \right] B_I^F
           +   \left[ 4 k_F^2 + q^2 - 4 \moqd \right] \mrho \left[ B_I^F \right]^2  \q , \\
\nn
\eqr

\bqr
\wwwuuuI & = & \wuuI - 2 q^2 \left( 4 B^{\Delta s}_I + B^T_I \right)  
                     + 8 q^4 \left[ C^{\nabla J}_{I} \right]^2 
                       \frac{\pbdbt}{1 + q^2 \pbdbt \wdzI} 
                     - 4 \moqd B_I^F                                                           \nn \\
         & + & \left[ B_I^F \right]^2
               \left\{ q^2 \mrho + \frac{1}{4} \left[ q^2 - 4 \moqd \right]^2 \chiz
             - 2 k_F^2 \left[ q^2 + 4 \moqd \right] \chid + 4 k_F^4 \chiq \right\}        \q .  \\
\nn
\eqr

\ei

\ewt

The $X^{{\rm (1,M,I)}}$ coefficients occurring in the previous expressions of the 
$S=1$ response functions are defined in the~\citeAppendix{app:wcoef}, while the momenta $\beta_{i}$ were already defined in Appendix D of article I.
 
%

\section{the $W_i^{(\alp)}$, $W_{i,L}^{(\alp)}$ and $X^{(\alp)}$ coefficients.}
\label{app:wcoef}

In order to simplify all the written formula in the presence of a tensor part
in the Skyrme interaction, the $W_1^{\text{(S,I)}}$ and
$W_2^{\text{(S,I)}}$ coefficients have been defined as

\bqr
\frac{1}{4}W^{(0,0)}_{1}&=&2A_{0}^{\rho 0}+(2+\gamma)(1+\gamma)A_{0}^{\rho \gamma}\rho^{\gamma}-\left[ 2C_{0}^{\Delta \rho}+\frac{1}{2}A_{0}^{\tau}\right]q^{2},\nonumber\\
\frac{1}{4}W^{(0,1)}_{1}&=&2A^{\rho 0}_{1}+2A^{\rho,\gamma}_{1}\rho^{\gamma} -\left[2 A_{1}^{\Delta \rho}+\frac{1}{2}A_{1}^{\tau}\right]q^{2},\nonumber\\
\frac{1}{4}W^{(1,0)}_{1}&=& 2A_{0}^{s,0}+2A_{0}^{s\gamma}\rho^{\gamma} -\left[2A_{0}^{\Delta s}+\frac{1}{2}A_{0}^{T} \right]q^{2},\nonumber\\
\frac{1}{4}W^{(1,1)}_{1}&=& 2A_{1}^{s,0}+2A_{1}^{s\gamma}\rho^{\gamma}- \left[2A_{1}^{\Delta s}+\frac{1}{2}A_{1}^{T} \right] q^{2},\nonumber\\
\frac{1}{4}W^{(0,0)}_{2}&=&A_{0}^{\tau}, \nonumber\\
\frac{1}{4}W^{(0,1)}_{2}&=&A_{1}^{\tau},\nonumber\\
\frac{1}{4}W^{(1,0)}_{2}&=&A_{0}^{T},\nonumber\\
\frac{1}{4}W^{(1,1)}_{2}&=&A_{1}^{T},\nn
\eqr

\noi for the residual interaction of a general Skyrme functional in the Landau limit (see Eq.(\ref{limitL})) we define the  $W_{1,L}^{\text{(S,I)}}$ and $W_{2,L}^{\text{(S,I)}}$ coefficients as 

\bqr
\frac{1}{4}W^{(0,0)}_{1,L}&=&2C_{0}^{\rho 0}+(2+\gamma)(1+\gamma)C_{0}^{\rho \gamma}\rho^{\gamma}, \nonumber\\
\frac{1}{4}W^{(0,1)}_{1,L}&=&2C^{\rho 0}_{1}+2C^{\rho,\gamma}_{1}\rho^{\gamma} ,\nonumber\\
\frac{1}{4}W^{(1,0)}_{1,L}&=& 2C_{0}^{s,0}+2C_{0}^{s\gamma}\rho^{\gamma},\nonumber\\
\frac{1}{4}W^{(1,1)}_{1,L}&=& 2C_{1}^{s,0}+2C_{1}^{s\gamma}\rho^{\gamma},\nonumber\\
\frac{1}{4}W^{(0,0)}_{2,L}&=&C_{0}^{\tau}, \nonumber\\
\frac{1}{4}W^{(0,1)}_{2,L}&=&C_{1}^{\tau},\nonumber\\
\frac{1}{4}W^{(1,0)}_{2,L}&=&C_{0}^{T},\nonumber\\
\frac{1}{4}W^{(1,1)}_{2,L}&=&C_{1}^{T},\nn
\eqr

\noi and the same way for the $X^{\rm (1,M,I)}$ coefficients can be written in terms of the EDF coupling constants as

\bqr
\label{definizioneX}
X^{{\rm (1,0,I)}}    & = & 8 \, q^2 \, \left[ B^F_I \right]^2                \nn \\
                & \times & 
\frac{\bdbt}{1 + q^2 \pbdbt \left[ W_{2}^{(\text{1,I})}+4B_{I}^{T} + 6 B_I^F \right]}    \, , \nn \\
X^{{\rm (1,\pm1,I)}} & = & 8 \, q^2 \, \left[ B_I^F \right]^2                \nn \\
                & \times & 
\frac{\bdbt}{1 + q^2 \pbdbt \left[ W_{2}^{(\text{1,I})}+4B_{I}^{T} \right]}              \, \nn.  \\
\nn
\eqr

%


\end{appendix}


\bibliography{linear_response_moments}


\end{document}